\documentclass[twocolumn,english,aps, pre]{revtex4-1}
\usepackage[T2A,LGR,T1]{fontenc}
\usepackage[latin9]{inputenc}
\usepackage{color}
\usepackage{units}
\usepackage{amsmath}
\usepackage{amssymb}
\usepackage{graphicx}
\usepackage{esint}

\makeatletter

\DeclareRobustCommand{\greektext}{%
  \fontencoding{LGR}\selectfont\def\encodingdefault{LGR}}
\DeclareRobustCommand{\textgreek}[1]{\leavevmode{\greektext #1}}
\ProvideTextCommand{\~}{LGR}[1]{\char126#1}

\DeclareRobustCommand{\cyrtext}{%
  \fontencoding{T2A}\selectfont\def\encodingdefault{T2A}}
\DeclareRobustCommand{\textcyr}[1]{\leavevmode{\cyrtext #1}}

\AtBeginDocument{
\usepackage{subfig}
\usepackage{subfloat}
\captionsetup[subfigure]{labelformat=simple}
\captionsetup[subfloat]{position=top,labelfont={large,bf},textfont=normalfont,singlelinecheck=off,justification=raggedright}

}

\@ifundefined{showcaptionsetup}{}{%
 \PassOptionsToPackage{caption=false}{subfig}}
\usepackage{subfig}
\makeatother

\usepackage{babel}
\begin{document}

\title{\textcolor{black}{Critical dynamics of gene networks is a mechanism
behind aging and Gompertz law}}

\author{Dmitriy Podolskiy$^{1,2}$, Ivan Molodtsov$^{3,4}$, Alexander Zenin$^{3}$,
Valeria Kogan$^{3}$, Leonid I. Menshikov$^{3}$, Vadim N. Gladyshev$^{2}$,
Robert J. Shmookler Reis$^{6,7,8}$ \& Peter O. Fedichev$^{3,5}$}

\affiliation{$^{1)}$Massachusetts Institute of Technology, 77 Massachusetts Ave.,
Cambridge, MA, 02139, USA,}

\affiliation{$^{2)}$Division of Genetics, Department of Medicine, Brigham and
Women\textquoteright s Hospital, Harvard Medical School, Boston, MA
02115, USA,}

\affiliation{$^{3)}$Gero LLC, Moscow, Russia. Novokuznetskaya 24/2,}

\affiliation{$^{4)}$Faculty of Bioengineering and Bioinformatics, Lomonosov Moscow
State University, Moscow 119234, Russian Federation,}

\affiliation{$^{5)}$Moscow Institute of Physics and Technology, 141700, Institutskii
per. 9, Dolgoprudny, Moscow Region, Russian Federation,}

\affiliation{$^{6)}$McClellan VA Medical Center, Central Arkansas Veterans Healthcare
System, Little Rock, AR, USA,}

\affiliation{$^{7)}$Department of Biochemistry and Molecular Biology, University
of Arkansas for Medical Sciences, Little Rock, AR, USA,}

\affiliation{$^{8)}$Department of Geriatrics, University of Arkansas for Medical
Sciences, Little Rock, AR, USA.}
\begin{abstract}
Although accumulation of molecular damage is suggested to be an important
molecular mechanism of aging, a quantitative link between the dynamics
of damage accumulation and mortality of species has so far remained
elusive. To address this question, we examine stability properties
of a generic gene regulatory network (GRN) and demonstrate that many
characteristics of aging and the associated population mortality rate
emerge as inherent properties of the critical dynamics of gene regulation
and metabolic levels. Based on the analysis of age-dependent changes
in gene-expression and metabolic profiles in \emph{Drosophila melanogaster},
we explicitly show that the underlying GRNs are nearly critical and
inherently unstable. This instability manifests itself as aging in
the form of distortion of gene expression and metabolic profiles with
age, and causes the characteristic increase in mortality rate with
age as described by a form of the Gompertz law. In addition, we explain
late-life mortality deceleration observed at very late ages for large
populations. We show that aging contains a stochastic component, related
to accumulation of regulatory errors in transcription/translation/metabolic
pathways due to imperfection of signaling cascades in the network
and of responses to environmental factors. We also establish that
there is a strong deterministic component, suggesting genetic control.
Since mortality in humans, where it is characterized best, is strongly
associated with the incidence of age-related diseases, our findings
support the idea that aging is the driving force behind the development
of chronic human diseases.
\end{abstract}
\maketitle

\subsection*{Introduction}

Aging is a complex biological process. It occurs at every possible
scale characterizing the living organism, ranging from the single
cell level (\emph{e.g.}, oxidative damage, somatic mutations) to the
level of interaction between different organs (\emph{e.g.}, failure
of individual organs with age or accumulation of dangerous byproducts
of metabolic activity such as arterial plaque). This is why it has
been notoriously difficult to pinpoint the ultimate cause, a single
molecular mechanism behind aging and relate it to such life-long consequences
as the onset and progression of age-related diseases and, finally,
death. As a consequence, a wide spectrum of propositions on the origin
of aging has emerged over the years such as theories that view aging
as a pre-programmed process \cite{Skulachev2001,Longo2005}, various
concepts involving damage accumulation \cite{Partridge2002,Sinclair2009,Gladyshev2012,Gladyshev2013,ORGEL1973},
hyperfunction theory \cite{blagosklonny2006aging,blagosklonny2009growth,DeMagalhaes2012},
disposable soma theory \cite{Kirkwood1977,Kirkwood2000}, antagonistic
pleiotropy \cite{williams2001pleiotropy} and mutation accumulation
theory \cite{medawar1952unsolved}. Since the theories of aging often
differ greatly from each other both in spirit and letter and reflect
upon different ``facets'' of aging \cite{reis1989model,Gladyshev2012,Gladyshev2013,gladyshev2016aging},
it is not always possible to see their mutual inconsistencies and
compatibilities.

One of the key ideas in the study of aging is the causal role of molecular
damage. Over 40 years ago Leslie Orgel proposed an attractive damage-based
theory\cite{ORGEL1973}, in which he argued that molecular corruption
of enzymatic machinery of transcription and/or translation would produce
positive feedback, leading to an \textquotedblleft error avalanche\textquotedblright .
We recently proposed a more general hypothesis of error accumulation\cite{kogan2014stability},
supported by a semi-quantitative stability assessment for a generic
gene regulatory network (GRN). We suggested a causal role of regulatory-error
accumulation in the age-dependent increase in mortality rate, reflecting
limited ability of cellular damage-response pathways to repair the
consequences of internal or external stresses However, no underlying
mechanism has been established thus far, for the interplay between
the dynamics of error accumulation in transcription/translation/metabolic
processes and the increase in mortality rates.

To identify such a link, here we develop a theory of GRN stochastic
critical dynamics and show that under very generic assumptions GRNs
of most species should be inherently unstable in order to be compatible
with age-dependent behaviour of mortality rates. Applying the method
of proper orthogonal decomposition \cite{Antoulas2009} to the publicly
available age-dependent transcriptome and metabolome datasets of \emph{Drosophila
melanogaster }\cite{pletcher2002genome,avanesov2014age}, we demonstrate
explicitly that autocorrelation functions of transcriptional and metabolic
datasets for \emph{D. melanogaster} exhibit stochastic exponential
instability with a characteristic time scale, $t_{\alpha}$, which
is of the same order as mortality rate doubling time $t_{{\rm MRDT}}$
of \emph{D. melanogaster}. We provide explicit solutions for the stochastic
aging dynamics of realistic GRNs to obtain expressions for the age-dependent
mortality increase. We argue that the two time scales, $t_{\alpha}$
and $t_{{\rm MRDT}}$, should coincide for any species in which mortality
follows the Gompertz equation. We are thus able to causally relate
GRN instability to the characteristic Gompertzian increase of the
mortality rate in populations. The exponential increase of mortality
eventually slows down and the mortality rate is expected to approach
a plateau, $M(t\agt t_{{\rm ls}})\sim1/t_{{\rm MRDT}}$, at late ages.
The result should be valid for sufficiently long lived species, $t_{{\rm ls}}\gg t_{\alpha}$.
The prediction is confirmed by quantative analysis of the transcriptomes,
metabolomes and mortality curves for populations of \emph{D. melanogaster}
and mortality curves for very large cohorts of medflies \cite{vaupel1998biodemographic}.
We establish specific metabolites and genes as biomarkers of aging
in\emph{ D. melanogaster}, examine the respective human orthologs,
and provide their associations with genes and pathways commonly related
to the major human chronic diseases. 

We show that the dynamics of aging, as manifested in transcriptional
and metabolic profiles, can be decomposed into both stochastic component,
related to regulatory error accumulation in transcription/translation/metabolic
pathways, and strong deterministic component, which can be naturally
associated with a genetic program of an organism. We show that the
presented model may serve to embrace most of the known features of
the aging processes and hence provides a novel and universal basis
for quantitative analysis of diverse age-dependent processes. We believe
that the theoretical ideas presented in the paper could be useful
in the identification of biomarkers and, after appropriate development,
in mechanistic studies of processes that regulate aging in any sufficiently
long-lived organism.

\section*{Results}

\subsection*{Critical dynamics of Gene Regulatory Networks: quantitative model
of aging}

In this Section, we first explain the theoretical basis of our quantitative
analysis of transcriptional and metabolic profile changes with age
and construct a quantitative model of aging. We consider the case
of the GRN described by a generic set of non-linear matrix equations
of systems biology
\[
g\left(x,dx/dt,d^{2}x/dt^{2},...\right)=F.
\]
 Here $g$ is a vector function of any number of quantities representing
the state of the system, $x$, and its time derivatives which characterize
the interactions between different components of $x$. If, for example,
the state vector $x$ consists of the gene expression levels only,\emph{
e.g.} as in \cite{pletcher2002genome}, the function $g$ may be thought
of as encoding pathways to which the given genes contribute. In addition
to gene expression levels, the state vector $x$ may include other
descriptors, such as levels of metabolites \cite{avanesov2014age}
or methylation levels in DNA \cite{horvath2013dna}. As the system
state vector $x$ is not observable in its entirety, we presume that
any sufficiently large part of it captures representative information
about ``macroscopic'' properties of the biological system state
at the organismal level. For the sake of simplicity, below we will
refer to gene expression only, if not stated otherwise. 

The vector $F$ describes the action of external or internal stress
factors affecting components of $x$. The vector $F$ may depend on
time, $t$: $F(t)=F_{0}+\delta F(t)$, where $F_{0}$ and $\delta F$
represent the mean stress and the fluctuations of stress levels, respectively.

Over long times, gene expression levels typically fluctuate near certain,
relatively slowly changing, mean values, corresponding to the average
homeostatic state of an organism. Therefore, we assume that there
is always a quasi-stationary point $x_{0}$ (homeostasis of the organism),
given by the solution of the stationary equation $f(x_{0})=F_{0}$.
As a result, the slow dynamics of the fluctuations of the gene expression
levels, $\delta x=x-x_{0}$, in its leading order obeys th\textcyr{\cyre}
matrix stochastic equation for a linear continuum-limit of a Markov
chain: 
\begin{equation}
D\delta\dot{x}+K\delta x=\delta F,\label{eq:mainEq1}
\end{equation}
where the matrices $D$ and $K$ describe the dynamical relaxation
properties and the interactions between the components of the gene
regulatory network, respectively (see Supplementary Information, Section
\ref{subsec:Slow-stochastic-dynamics-1}, for in-depth discussion
of Eq.(\ref{eq:mainEq1}) derivation). 

It has been previously suggested that, in most species, GRNs operate
close to a stability-instability transition or order-disorder bifurcation
\cite{balleza2008critical,krotov2014morphogenesis,kogan2014stability}.
The transition from stability to instability in networks with the
network graphs not possessing specific symmetries is typically associated
with occurrence of co-dimension $1$ bifurcations \cite{Fiedler2002,Seydel2009}.
Such transitions are characterized by the loss of stability along
a single direction in the state vector space, corresponding to the
first principal component, while contributions from all other principal
components remain stable. This situation, known in the literature
as a saddle-node bifurcation \cite{Fiedler2002}, is realized when
the smallest real eigenvalue, $\epsilon$, of the matrix $K$ tends
to zero, $\epsilon\rightarrow0,$ and then becomes negative. Accordingly,
the stationary solution for $x_{0}$ ceases to exist, and the homeostatic
state of the organism starts to change in time. As we shall argue,
this very instability is directly responsible for the process of aging. 

\begin{figure*}
\subfloat[]{\includegraphics[width=0.5\textwidth]{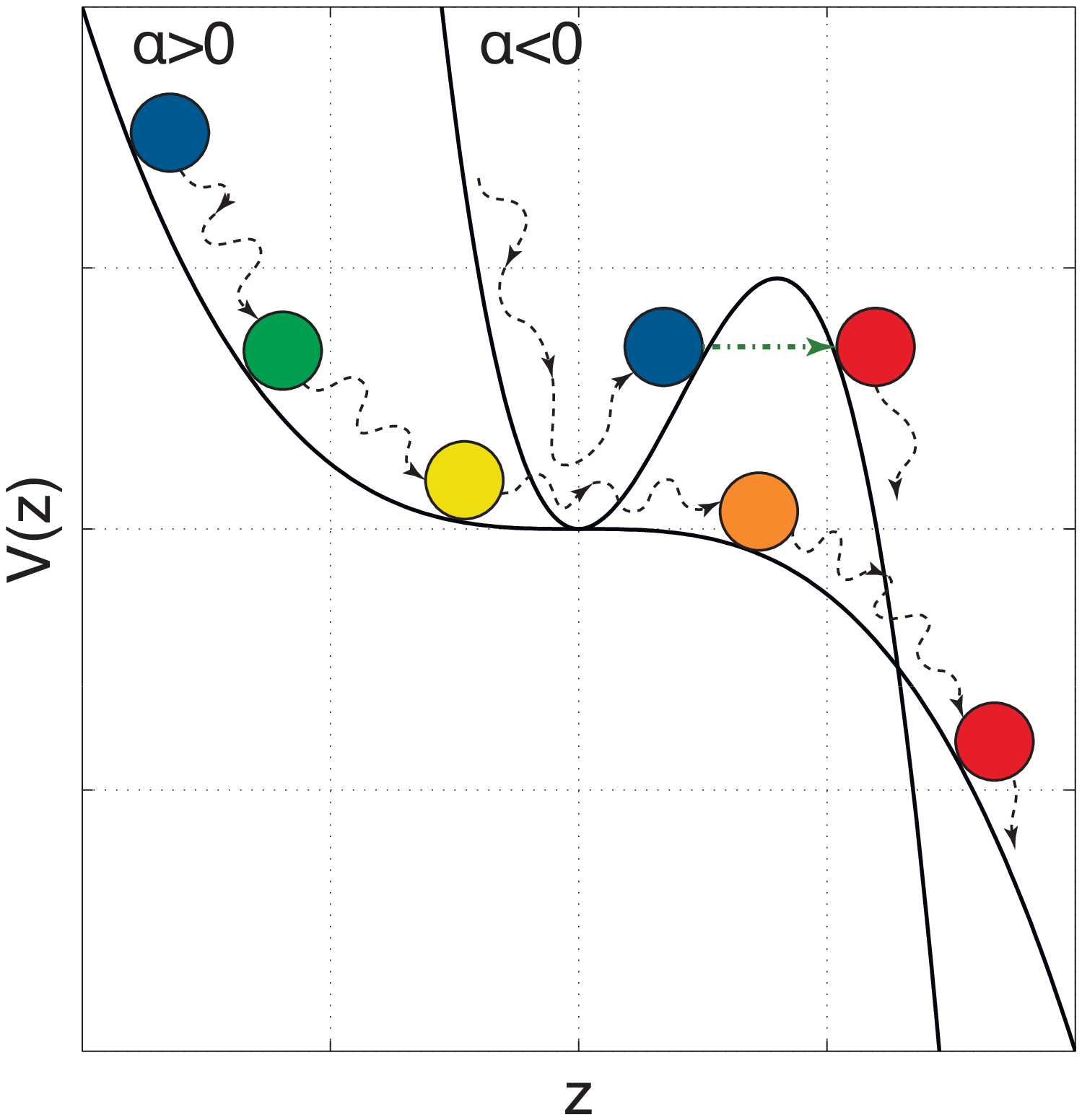}}\subfloat[]{\includegraphics[width=0.5\textwidth]{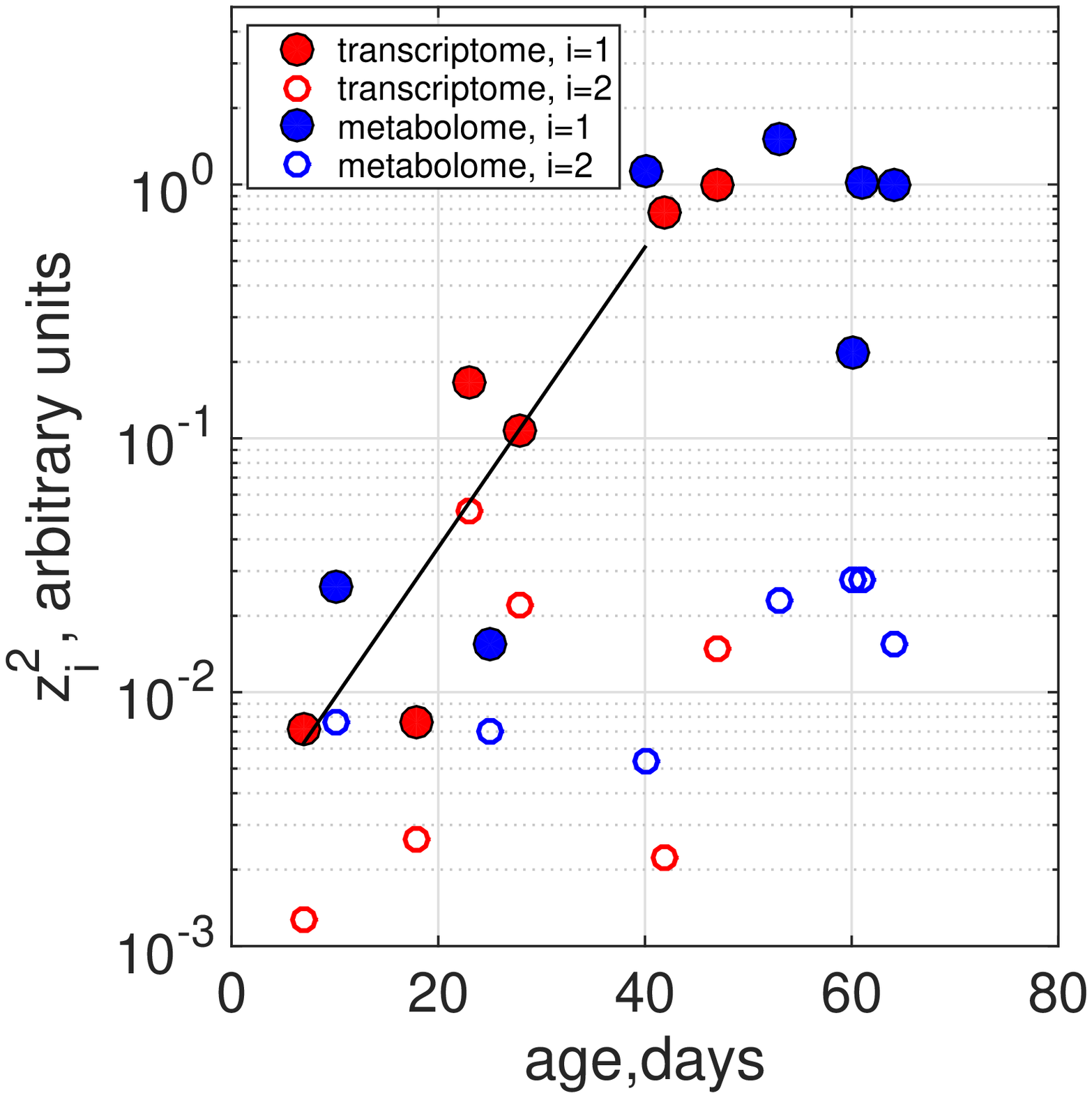}}

\caption{(A) Schematic representation of the stochastic dynamics in the effective
potential $V(z)$ corresponding to the cases of unstable ($\alpha>0$)
and stable ($\alpha<0$) GRNs, see Eq.(\ref{eq:1DoF}). (B) Age dependence
of $z_{1,2}=(\delta x^{T}\cdot b_{1,2})$ of the expressome state
projections onto the vectors $b_{1,2}$ corresponding to the first
two principle components as functions of age (gene expressions for
normally fed \textit{D. melanogaster} from \cite{pletcher2002genome},
red, and metabolite levels for normally fed \textit{D. melanogaster}
from \cite{avanesov2014age}, blue). Here $E(...)$ stands for the
average over the biological repeats; $(a^{T}\cdot b)$ denotes a scalar
product of two vectors $a$ and $b$. \label{fig:SchematicVandPCA}}
\end{figure*}

\begin{figure*}
\subfloat[]{\includegraphics[width=0.5\textwidth]{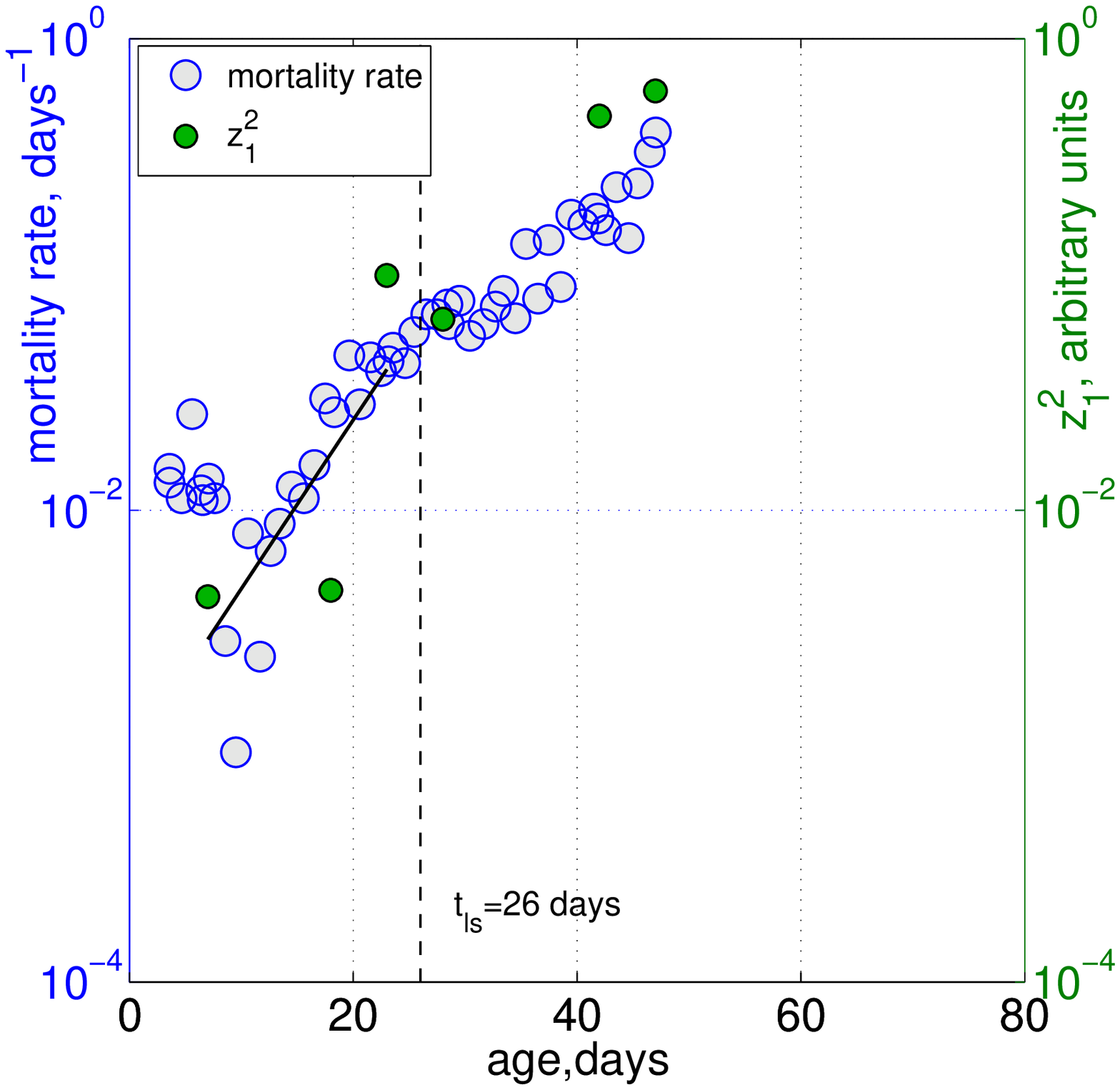}
}\subfloat[]{\includegraphics[width=0.5\textwidth]{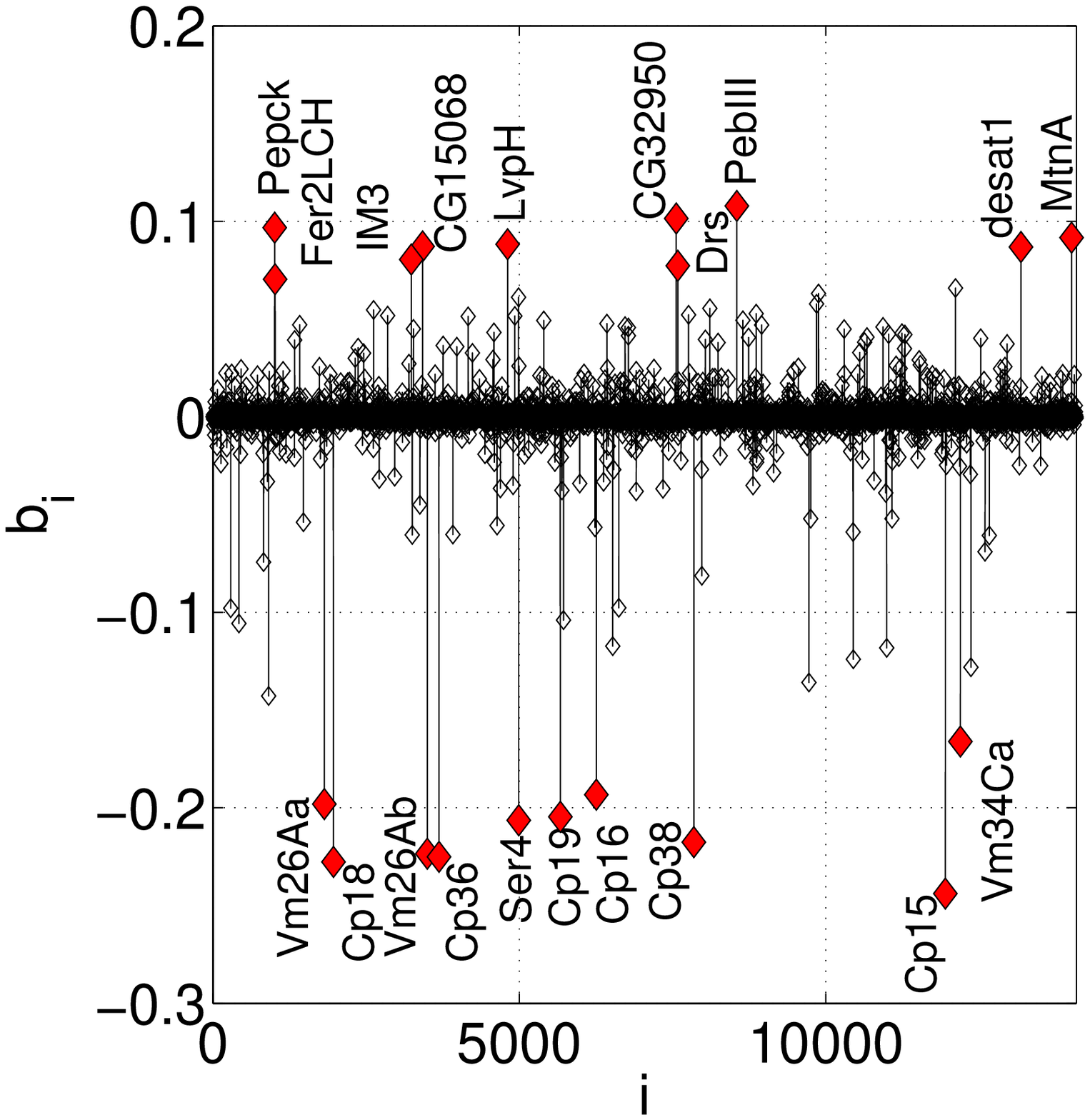}}

\subfloat[]{\includegraphics[width=0.5\textwidth]{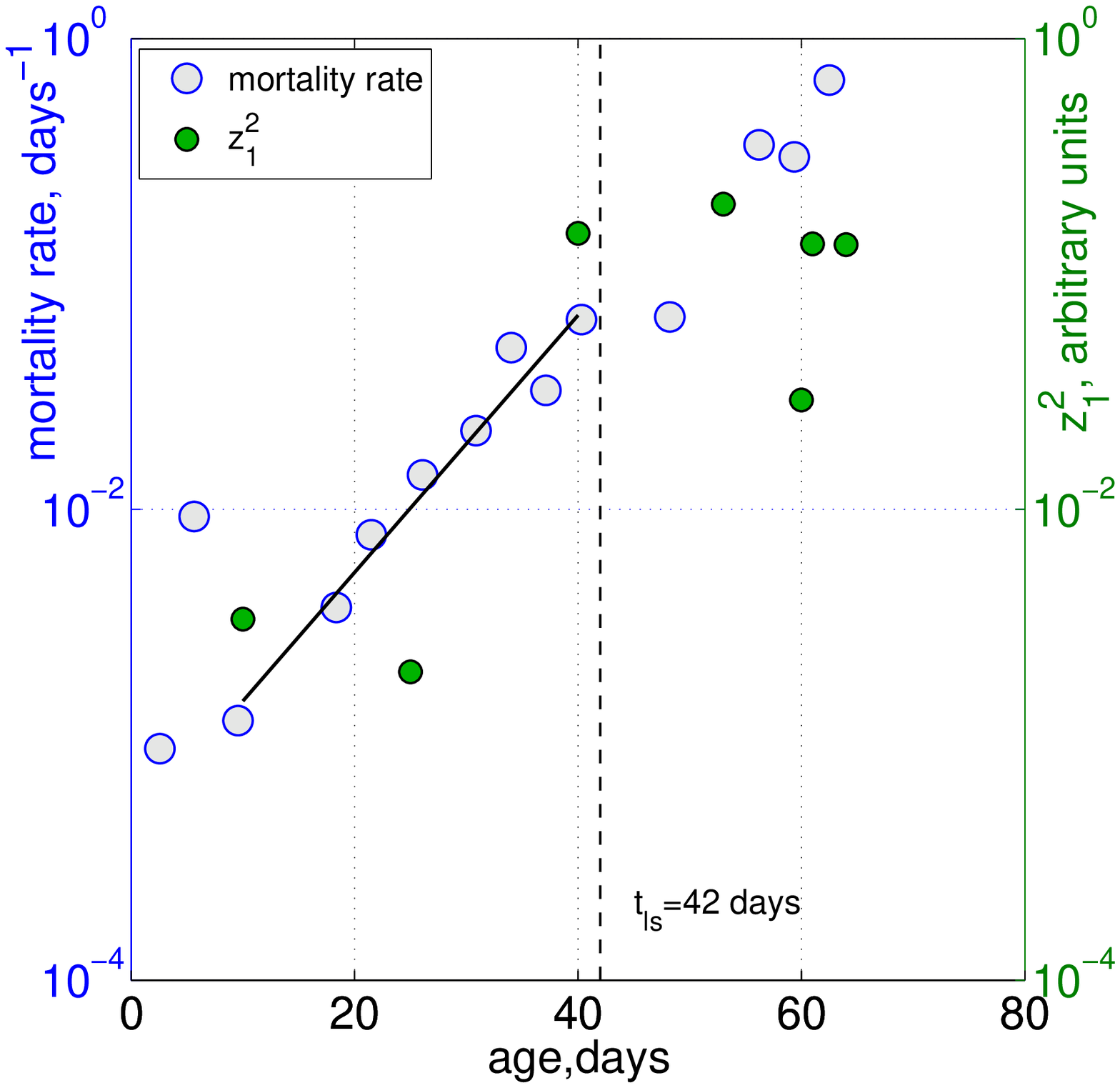}}\subfloat[]{\includegraphics[width=0.5\textwidth]{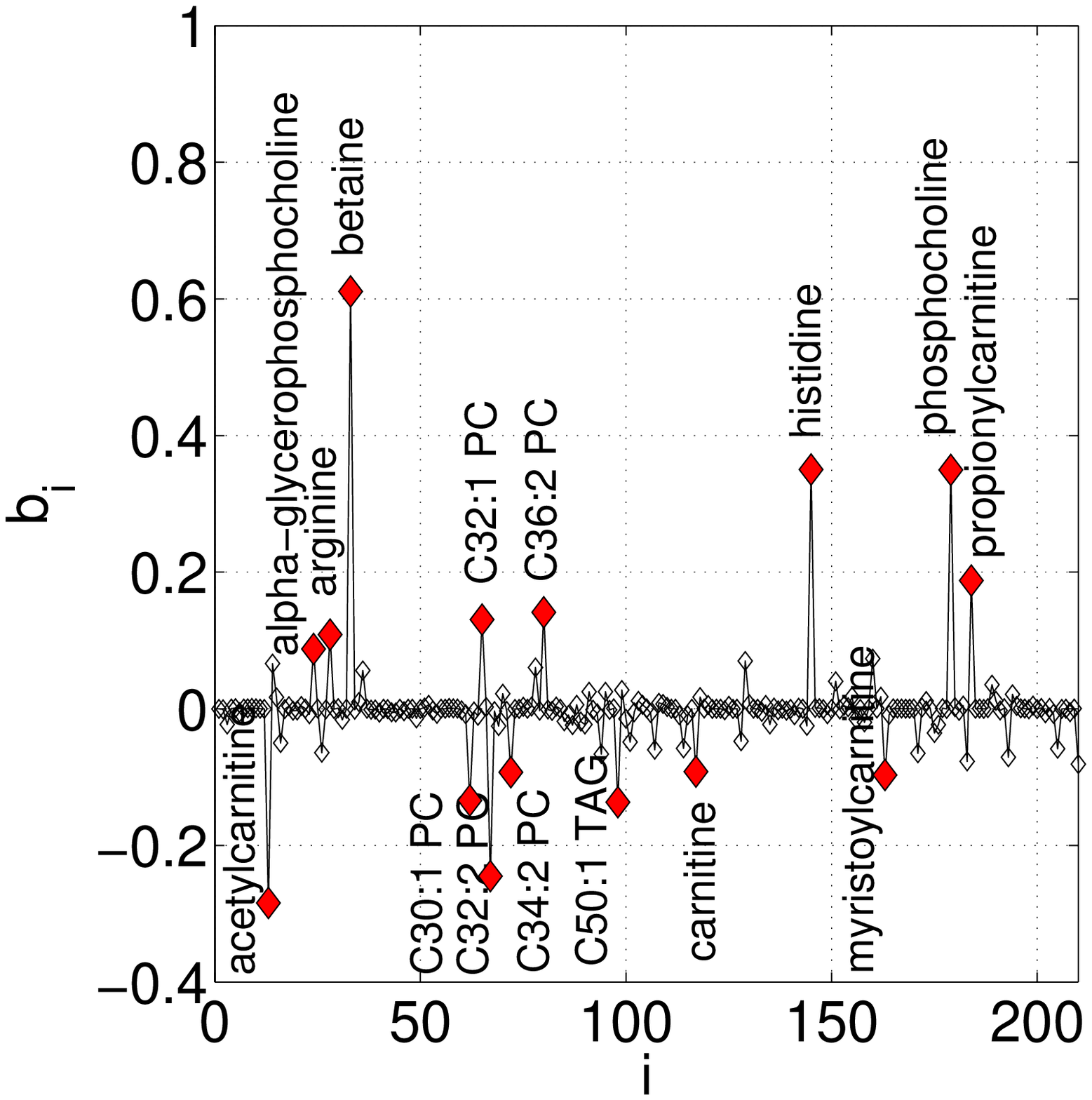}}

\caption{Basic biological evidence for the presented stochastic model of aging:
(A) Mortality rate along with the projection $z$ of the system state
vector, $\delta x$, represented by the transcriptomes of aging\emph{
D. melanogaster} \cite{pletcher2002genome}, onto the singular direction
$b$ of the covariance matrix $E(\delta x(t)\delta x^{T}(t'))$ as
a function of age. (B) Components of the vector $b$ for \emph{D.
melanogaster}. Peaks correspond to the values of the components of
the vector $b$ for every gene in the dataset. (C) Mortality rate
and the projection $z$ of the system state vector $\delta x$, represented
by the metabolomes of\emph{ D. melanogaster} as a function of age
\cite{avanesov2014age}; (D) Components of the vector \textbf{$b$}
for \emph{D. melanogaster}, computed using only the levels of targeted
metabolites \cite{avanesov2014age}. \label{fig:Main-TRMET}}
\end{figure*}

\begin{figure*}
\includegraphics[width=0.9\textwidth]{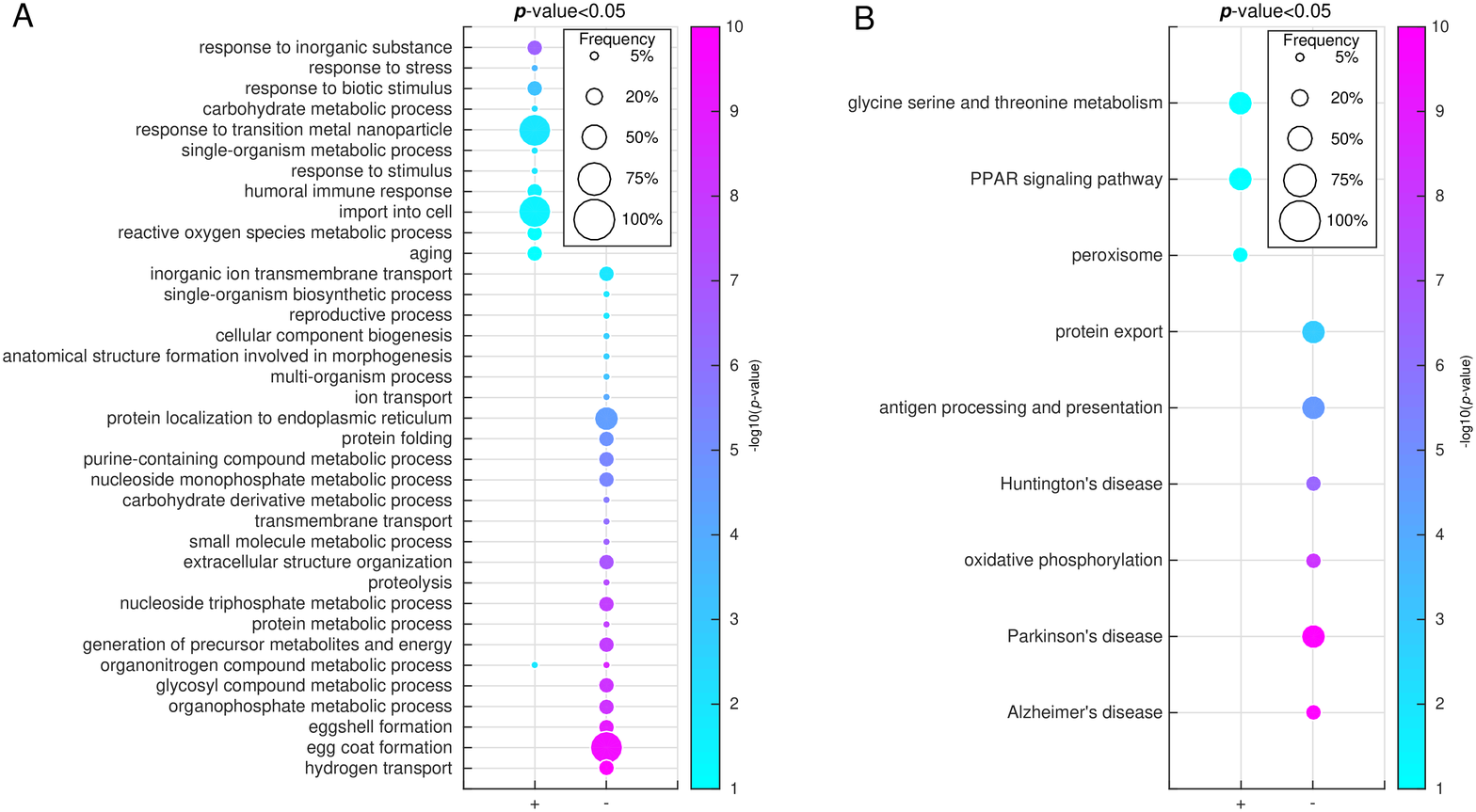}

\caption{Biological characterization of the ``leading direction of aging''
$b$ in \emph{D. melanogaster}: (A) Gene Ontologies (Biological Process
only) enrichment by the leading positive and negative components of
the vector $b$ computed using gene expression transcripts; (B) KEGG
pathways enrichment for human orthologs of the gene sets representing
the leading components of $b$, corresponding to positively and negatively
regulated genes. \label{fig:GOntologizer}}
\end{figure*}

To be specific, hereinafter we focus on this scenario. As explained
in Supplementary Information, Appendix \ref{subsec:Slow-stochastic-dynamics-1},
the derivative $\delta x$ of any gene expression level becomes critical
close to the co-dimension 1 bifurcation point. This implies that fluctuations
of the expressome $\delta x$ are strongly amplified with age, the
phenomenon known as critical slowing-down (see \emph{e.g.} \cite{Suzuki2009}).
In particular, the autocorrelation time scale $t_{{\rm auto}}=(a^{T}\cdot Db)/\epsilon$
of the expressome diverges as $\epsilon\to0$, indicating that stochastic
dynamics of $\delta x$ is very slow near the point of bifurcation.
The fluctuations of the expressome state vector $\delta x$ including
those created in response to persistent external stress, are mostly
amplified along the direction $b$ of the right eigenvector of matrix
$K$ corresponding to the vanishing eigenvalue $\epsilon$. The latter
observation is supported by the analysis of experiments in \emph{D.
melanogaster}, where transcriptional responses to different stress
factors were found to contain a good fraction of similar differentially
expressed genes in the limit of both nearly lethal \cite{brown2014diversity}
and weak stressors \cite{10.1371/journal.pone.0086051}. 

Accordingly, over times comparable to the average lifespan, the gene
expression fluctuations are dominated by transcript-level changes
along the vector corresponding to the leading principal component,
which is collinear with the direction of the vector $b$. We conclude
that for long time intervals evolution of the vector of the GRN state
vector, $\delta x$, can be accurately described using a single variable
$z$, such that $\delta x\approx z\cdot b$. Accordingly, Eq.(\ref{eq:mainEq1})
can be reduced to the equation 
\begin{equation}
\frac{dz}{dt}+\frac{\partial V(z)}{\partial z}=f,\label{eq:1DoF}
\end{equation}
describing the Brownian motion of a particle in an effective potential
$V(z)\approx-\alpha z^{2}/2$, for sufficiently small $z$, and the
quantity $\alpha=-\epsilon/(a^{T}\cdot Db)$ characterizes the ``stiffness''
of the gene regulatory network. The stochastic force, $f=(a^{T}\cdot\delta F)/(a^{T}\cdot Db)$,
is a single unified quantitative measure of stochastic component of
all external and internal stress, and is characterized by the diffusion
coefficient $\Delta$ (see Supplementary Information, Appendix \ref{subsec:Slow-stochastic-dynamics-1}
for the in-depth derivation and necessary justification).

Depending on the specific morphological properties of the gene regulatory
network, the curvature of $V(z)$ at small $z$ can be negative, $\alpha<0$,
or positive, $\alpha>0$, as shown in Figure \ref{fig:SchematicVandPCA}A.
The first case will be briefly considered in Discussion section. In
the latter case, the GRN is inherently unstable, and the variance
of gene expression levels, $C(t)$, computed using the solutions of
Eq.(\ref{eq:1DoF}), grows exponentially with age as

\begin{equation}
C(t)=E(\delta x(t)\delta x^{T}(t))\approx E(\delta x_{0}^{2})b\cdot b^{T}\exp(2\alpha t).\label{eq:CIJt}
\end{equation}
The expression (\ref{eq:CIJt}) is fully consistent with earlier observation
of increasing biological variability with age \cite{variability,variabilityGerm}.

Criticality of gene expression-level dynamics criticality implies
that the parameter $\alpha$ is close to zero and hence the expressome
covariance matrix $C(t)$ should be a matrix with special structure.
Eq.(\ref{eq:CIJt}) implies that in this case the right eigenvector
of the covariance matrix corresponding to its largest eigenvalue coincide
with the right eigenvector $b$ of the matrix $K$ corresponding to
vanishing eigenvalue $\epsilon\rightarrow0$ (see Supplementary Information,
Appendix \ref{subsec:Slow-stochastic-dynamics-1}). What is more,
the largest eigenvalue should be much larger than all other eigenvalues
of this matrix. In reality the properties of the experimental sample
covariance matrix may be degraded by the experimental noise and necessarily
very limited number of samples compared to the number of transcripts
or metabolites observed. 

These conclusions are supported by Principal Component Analysis (PCA)
of the age-dependent shifts of RNA-transcript and metabolite levels
in normally fed\textit{ D. melanogaster} from \cite{pletcher2002genome}
and \cite{avanesov2014age}, respectively. The covariance matrices
computed from such datasets are indeed almost singular, see Supplementary
Information, Figure \ref{fig:Five-largest-SVD}. Correspondingly,
the variation of the state vector $\delta x$ with age is dominated
by the change along the loading vector corresponding to the first
principal component, and $z\equiv z_{1}$, where $z_{1}=(\delta x^{T}\cdot b)$
is the first principal component, as described in Figure \ref{fig:SchematicVandPCA}B.
The variance along this vector grows very quickly with age, both in
the metabolome and the transcriptome, in accordance with Eq.(\ref{eq:CIJt}).
The exponent, roughly inferred from each of the datasets for ages
less than the mean lifespan, appears to be nearly the same, which
is an indication of the same GRN instability observed in each of the
experiments. For the sake of comparison, we have also plotted projections
of gene expression on the next principal component. It is easy to
see that the transcriptome and metabolome remain dynamically stable
along the direction of the second principal component. 

We thus hypothesize that at the origin, $z\approx0$, the expressome
state corresponds to a healthy or ``youthful'' state, while larger
values of $z$ phenotypically describe distorted states of aged animals.
Accordingly, aging manifests itself as a slow dynamics of expression
levels, an exponential roll-out along the singular, or ``aging''
direction $b$, from the younger animals to the older ones. The singular
``direction of aging'' $b$ is associated with a response to a generic
stress and determined by the properties of the interactions in the
network only. At least within the model assumptions, the direction
$b$ does not depend on the detrimental stress factors and appears
to be hard-coded in the genome, and hence the development along $b$
can be considered as an organism-level manifestation of an aging program
or quasi-program \cite{khalyavkin2014aging,arlia2014quasi,blagosklonny2012answering,blagosklonny2013mtor}).
The ``direction of aging'', vector $b$, is similar but not the
same as the differential expression vector in aging animals, since
the latter may include a contribution of faster modes represented
by other eigenvectors of the GRN connectivity matrix $K$ corresponding
to eigenvalues with larger absolute value. Since the network state
fluctuations along all the other directions are not related to aging
and are relatively small, we conclude that the deterministic component
of aging in $\delta x(t)$ already dominates over the stochastic one,
even early in life (also see below).

\subsection*{Gompertz mortality law and biological age}

To relate the GRN instability to the process of aging we further assume
that for every gene in the genome there exists a particular threshold
associated with its over- or under-expression, which leads to a strong
build-up of deleterious changes, which in it turn ultimately lead
to death. Since fluctuation of the expression levels can be written
in the form $\delta x\approx z\cdot b$, such deleterious thresholds
can be defined by the specific values of $z=Z_{i}$ for every gene,
which may be either positive or negative. It is reasonable to assume
that the death of a cell (and ultimately of an organism) occurs once
$z$ reaches the smallest (by absolute value) of the threshold values,
$z=Z$. In Supplementary Information Section \ref{subsec:gene-network-kinetics}
we argue, that as the expressome state deviates from the initial point,
the higher order non-linearities in Eq.(\ref{eq:mainEq1}) and therefore
in the effective potential $V(z)$ of Eq.(\ref{eq:1DoF}) can no longer
be neglected. As expected from the argument, we are able to show in
Supplementary Information section \ref{subsec:gene-network-kinetics},
that the threshold value $Z$ can be equally well introduced formally
in the model as the single effective property of the non-linearity
in such a way that $Z$ is large whenever the non-linear interactions
are weak.. 

Initially, the probability density along the aging direction is localized
at a small value of $z$ and the mortality is also very small. If
severe deleterious changes occur late in life, i.e. if $\gamma=\alpha Z^{2}/\Delta\gg1$,
the lifespan greatly exceeds $t_{\alpha}$, the non-linearity is small,
and the mortality increases exponentially with age first, then the
growth decelerates, and, eventually, the mortality reaches a plateau.
At intermediate ages, in a narrow range close to the average lifespan
of the animals, the mortality can be approximated with
\begin{equation}
M\approx\text{0.61\ensuremath{\alpha}}e^{\alpha(t-t_{{\rm ls}})}\sim\frac{\Delta}{Z^{2}}e^{\alpha t}.\label{eq:GompertzMortality}
\end{equation}
This is a form of the Gompertz law \cite{gompertz1825nature} (see
Supplementary Information, Appendix \ref{subsec:gene-network-kinetics}
for the complete analytical solution for the age-dependent mortality
in a closed form for all ages). The Gompertz exponent $\alpha$ is
related to the Mortality Rate Doubling Time (MRDT), $t_{{\rm MRDT}}$,
by equation $\alpha=\log(2)/t_{{\rm MRDT}}$. In practice, the value
of $\alpha$ is often determined from the survival records at the
age, corresponding to a narrow interval of width $\sim1/\alpha$ around
the mean lifespan, when most of the animals die. Therefore Eq.(\ref{eq:GompertzMortality})
suggests that the empirical Gompertz exponent should be very close
to the GRN ``stiffness''. 

On the other hand, the Initial Mortality Rate (IMR), $\Lambda_{{\rm IMR}}$,
is related both to the stress levels and to the time required to reach
the boundary $Z$ under the action of the random damaging forces only,
$\Lambda_{{\rm IMR}}=\Delta/Z^{2}$. Normally, the stochastic forces
are weak, $\gamma\sim(\alpha/\Lambda_{{\rm IMR}})$ is large, and,
therefore, average lifespan, $E(t_{{\rm ls}})\sim\alpha{}^{-1}\log\gamma\gg\alpha^{-1}$,
is also large compared to the characteristic time, associated with
the gene network instability, $t_{\alpha}\sim\alpha^{-1}$. Remarkably,
the lifespan in the model depends on the morphological properties
of the GRN, such as its topology and connectivity, through the matrix
$K$, and, more specifically, through its lowest eigenvalue $\epsilon\sim\alpha$.
The dependence of species lifespan on damaging factors through the
parameter $\Delta$ is, on the contrary, logarithmically weak. This
has been recently shown in analysis of data from \cite{katzenberger2013drosophila}
in our work\cite{kogan2014stability}, where long-term effects of
traumatic brain injury at young age led to significant, but small
changes in organismal lifespan. This also means that the gene-network
evolution along the aging direction is practically deterministic in
sufficiently long-lived organisms. At the same time, the stochastic
forces are still very important early in life, yielding a relatively
wide distribution of transcription profiles even in a group of otherwise
identical animals at birth. Our analysis proposes that the stochastic
component of the GRN dynamics explains the distribution of the observed
ages at death in such a cohort. 

Mortality in a form similar to Eq.(\ref{eq:GompertzMortality}) can
be associated with gene network instability in a simple phenomenological
model, where both IMR and MRDT were expressed in terms of generic
network parameters, such as translation, gene repair and protein turnover
rates, along with the stress levels, whereas mortality is directly
associated with the increasing number of regulatory errors \cite{kogan2014stability}.
Therefore, the quantitative indicator of the expressome remodeling
along the aging direction, $z$, can be interpreted as a measure of
damage accumulated over the lifespan of an organism. 

As we show in Supplementary Information, Appendix \ref{subsec:gene-network-kinetics},
the dynamics of gene expression levels $\delta x$ in the weak non-linearity,
or ``Gompertzian'' ,limit, $\gamma\gg1$, proceeds along a well-defined
trajectory, corresponding to a continuation of the development program.
The ``distance traveled'' along the aging direction up to current
moment of time, $z\sim(\delta x^{T}\cdot b)$, is, according to Eq.(\ref{eq:GompertzMortality}),
directly related to mortality and hence by itself is a good biomarker
of mortality and aging. These observations pave the way to define
biological ``clocks'' using any number of variables characterizing
a GRN state. For example, Figures \ref{fig:Main-TRMET}A \& C suggest
the possibility of developing transcriptome- and metabolome-derived
biological clocks for \emph{D. melanogaster}. Those clocks in turn
may be compared with the biomarkers of aging and the biological age
calculator (biological clock) similar to, for example, the one derived
from a regression model established using age-dependent DNA-methylation
patterns \cite{horvath2013dna}.

\subsection*{Association between GRN instability and mortality rate}

Our analysis of \emph{D. melanogaster} age-dependent transcriptional
data from \cite{pletcher2002genome} indicates that the gene expression
variance grows exponentially with age in agreement with Eq.(\ref{eq:CIJt}),
with $\alpha\approx19.1\,{\rm yr}^{-1}$, as shown in Figure \ref{fig:Main-TRMET}A.
In a similar fashion we have studied the ``direction of aging''
\textbf{$b$} in the metabolome of \emph{D. melanogaster} \cite{avanesov2014age}.
The projection of the metabolome state vector onto the singular direction
$b$ increases very quickly with age, along with the reported mortality
as shown in Figure \ref{fig:Main-TRMET}C. The mean lifespan of \emph{D.
melanogaster} is about 25 and $40$ days in both experiments, respectively,
and in the two cases the initial exponential growth of the projection
$z^{2}(t)$ ceases at $t\agt t_{ls}$, as predicted by the theory.
Figure \ref{fig:SchematicVandPCA}B represents a side-by-side comparison
of the variance computed from the gene expression and the metabolome
datasets. The values of instability rates, $\alpha$, recovered from
analysis of both experiments, are very similar and close to the value
of the Gompertz exponent $\alpha=\ln2/t_{{\rm MRDT}}=17.1\,{\rm yr}^{-1}$
corresponding to the mortality rate doubling time $t_{{\rm MRDT}}\approx0.04\,{\rm yr}$
\cite{tacutu2013human}. 

\subsection*{The pace of mortality-rate increase slows at advanced ages}

At more advanced ages, $t\agt E(t_{{\rm ls}})$, where $t_{{\rm ls}}$
is the average lifespan of the population, solutions of Eq.(\ref{eq:1DoF})
exhibit deceleration of mortality. More specifically, as shown in
\textcolor{black}{Supplementary Information, Appendix} \ref{subsec:gene-network-kinetics},
we expect that the mortality rate ceases growing and saturates at
a constant value, $M_{\infty}\sim\alpha$. This quantitative prediction
can be compared with the experimental data for medflies \cite{vaupel1998biodemographic},
where the cohort sizes were sufficient to observe both the regime
of exponentially increasing mortality and the mortality plateau (see
Supplementary Information, Figure \ref{fig:The-mortality-rate-medflies}).
For the male medflies we find the asymptotic value $M_{\infty}\approx0.16\,{\rm day^{-1}}$,
which is close to the Gompertz exponent $\alpha\approx0.22\,{\rm day}^{-1}$,
in accordance with theoretical prediction. The asymptotic value of
the mortality rate for female medflies, $M_{\infty}\approx0.12\,{\rm day}^{-1}$,
is also of the right order of magnitude, though nearly two times smaller
than the slope of the mortality exponent $\alpha=0.25\,{\rm day^{-1}}$.
Nevertheless, there is an approximate agreement between between the
asymptotic value $M_{\infty}$ of the mortality rate at late ages
and the value of the Gompertz exponent $\alpha$ in both cases. We
also note that the approximate Eq.(\ref{eq:GompertzMortality}) should
work better for longer living, or ``Gompertzian'', \emph{$\Lambda_{{\rm IMR}}\ll\alpha$,}
organisms.

\subsection*{Biomarkers of aging in\emph{ D. melanogaster}}

We computed the aging direction based on the transcriptomic data from
\emph{Drosophila melanogaster} \cite{pletcher2002genome}. Gene Ontolology
(GO) analysis of the leading components of $b$ vector is summarized
in Figure \ref{fig:GOntologizer}A, for Biological Process (BP) categories
only. The genes providing the most significant negative contribution
to the aging direction $b$ can be roughly split into three groups.
First, there is a large group of genes related to metabolic processes:
organophosphate metabolism, generation of precursor metabolites and
energy, membrane and mitochondrial ion transport. A detailed review
of the GO categories involved seems to indicate that most of the down-regulated
processes are related to oxidative phosphorylation or respiration.
Another large group of genes involves developmental processes, such
as anatomical structure formation involved in morphogenesis and extracellular
structure organization. Genes encoding eggshell and egg coat formation
correspond to strongly negative components of the vector $b$, which
points to the reproductive senescence in female flies. Finally, we
observe decline in processes related to proteostasis, including protein
folding, ER localization and proteolysis. 

As expected, the cluster of genes corresponding to the strongest positive
components of the vector $b$ includes genes associated with a generic
stress response. Among other leading components there are genes corresponding
to the innate immune response (which is also a stress response), in
agreement with previous conclusions \cite{khan2013aging}, and the
components of oxidation-reduction and reactive oxygen species metabolic
processes. 

We found that several transcription factors (TF) appear to be strongly
associated with aging. One of the top positively associated genes,
$hairy$, is a transcriptional suppressor, and believed to be a metabolic
switch, found up-regulated in the gene expression datasets characterizing
oxygen-deprived flies. Conversely, mutations in $hairy$ significantly
reduce hypoxia tolerance \cite{zhou2008mechanisms}. The only negatively
associated transcription factor, $eve$, is a transcription factor
involved in multiple organ and system development.

\subsection*{Human orthologs for biomarkers of aging in \emph{D. melanogaster}}

To obtain a better insight into aging of \emph{D. melanogaster} we
annotated gene expression changes involved in aging in flies with
human orthologs. This is possible to do since according to OrthoDB,
approximately 80\% of all genes in \textsl{D. melanogaster} have human
orthologs. We searched for human orthologs of the leading contributors
to the aging direction vector $b$ computed from the transcriptome
of \emph{D. melanogaster}. The results of this analysis are represented
in Figure \ref{fig:GOntologizer}B. Remarkably we observe transcriptional
changes associated with human age-related diseases: the downregulated
genes were enriched in KEGG categories corresponding to age-related
neurodegenerative diseases (Huntington\textquoteright s, Parkinson\textquoteright s
and Alzheimer\textquoteright s), and also the processes of cardiac
muscle contraction. We also observed that the downregulated genes
enriched in KEGG pathways were associated with oxidative phosphorylation
(consistent with the GO annotation and KEGG enrichment results above).
In contrast, KEGG PPAR signaling pathway is activated with age, which
may be another indication of the activation of anti-oxidant metabolism
systems as noted above. There is evidence suggesting that PPAR\textgreek{a}
and the genes under its control play a role in the evolution of oxidative
stress increases observed in the course of aging in mice \cite{poynter1998peroxisome}. 

Further evidence relating the stochastic instability of the gene regulatory
network to the common age-related diseases is a strong positive contribution
of the \textsl{Pepck} gene, which encodes a rate-controlling enzyme
of gluconeogenesis, the process by which cells synthesize glucose
from metabolic precursors. RNAi silencing of the Pepck gene was found
to be an effective way to reverse diabetes-induced hyperglycemia in
mice\cite{gomez2006overcoming}. It has also been shown that expression
of the Pepck ortholog in Caenorhabditis elegans correlates almost
perfectly with longevity in an isogenic series of longevity mutants\cite{tazearslan2009positive}.
Also, a transgenic mouse strain constitutively overexpressing muscle
PEPCK is muscular and long-lived \cite{hanson2008born}.

\section*{Discussion}

Our analysis suggests that aging involves an organismal level manifestation
of the inherent instability of gene regulatory networks. Proper orthogonal
decomposition of age-dependent transcriptional and metabolite profiles
shows that, for species which age in accordan\textcyr{\cyrs}e with
the Gompertz equation, the life-long dynamics of the transcriptional
and metabolite profiles can be effectively described in terms of stochastic
critical dynamics with a single degree of freedom; that dimension/component
is identified by the projection of the expressome on the right eigenvector
$b$ of the GRN connectivity matrix $K$, and associated with the
eigenvalue $\epsilon$ which vanishes at the critical point. We establish
that depending on the sign of $\epsilon,$ a GRN can be either stable
or unstable. We further propose that changes in metabolite levels
contribute to aging similarly to changes in GRN, e.g., through a buildup
of certain metabolites and molecular damage beyond a certain level.
We note that appearance of such deleterious thresholds is guaranteed,
if non-linearities in the equation (\ref{eq:mainEq1}) for expressome
levels are accounted for (see Supplementary Information, Appendix
\ref{subsec:gene-network-kinetics}). We show that in this limit mortality
follows the form of the Gompertz law, i.e. it first increases exponentially
with age and then saturates at a constant level, a prediction supported
by experimental evidence. As predicted, we find approximate agreement
between the rate of gene regulatory network instability and the value
of mortality rate saturation. Thus we are able to demonstrate that
Gompertzian aging is a property of organisms with inherently unstable
gene regulatory networks, and thus the gene network instability is
a process that directly contributes to aging, which can be observed
in the form of an exponential increase of all-cause mortality with
age.

Though stochasticity of the expressome is an essential part of the
model, it should be noted that in the Gompertzian limit a wide variety
of organisms exhibits similar dynamics of expression profiles. According
to our theory, genetically programmed connectivity matrix $K$ can
be represented by its low-rank approximation using its right eigenvector
$b$ and the corresponding eigenvalue $\epsilon$. Such a direction
can be interpreted as a genetically programmed mechanism of aging.
It is important to note that this direction is identifiable from experimental
data, thus allowing quantitative predictions to be made about a mechanism
behind aging.

The stochastic model of aging presented here is fully compatible with,
and in fact embraces, several prevailing theories and hypotheses of
aging, in the following sense:
\begin{itemize}
\item \textbf{programmed and quasi-programmed aging} \cite{Skulachev2001,Longo2005,DeMagalhaes2012}.
We have shown that aging-driven dynamics of the expressome $\delta x(t)$
can be separated into stochastic and deterministic components. The
deterministic component of $\delta x(t)$ starts to dominate over
the stochastic one even early in life, at roughly the same time as
the low-rank approximation of the GRN connectivity matrix $K$ becomes
prominent, as can be seen from the fact that the covariance matrix
is singular and dominated by the aging direction. As the dynamics
of the expressome $\delta x(t)$ is largely determined by the properties
of the matrix $K$, we interpret the low rank approximation of $K$
as a program or a quasi-program of aging \cite{blagosklonny2006aging},
though no statement about its evolutionary nature can be made. We
speculate that the same direction that corresponds to aging occurs
and is active during development, thus being consistent with quasi-programmed
aging theory.
\item \textbf{hyperfunction theory} \textbf{\cite{blagosklonny2006aging,blagosklonny2009growth}}.
The single mode $\delta x=z\cdot b$, describing a large cluster of
genes that make a dominant contribution to the process of aging, starts
dominating the aging dynamics soon after the completion of development,
and the stochastic effects on the expressome dynamics can be considered
relatively weak \cite{blagosklonny2006aging,blagosklonny2009growth}.
The expression of positive leaders of the aging direction $b$ will
grow with time, which can be interpreted as hyperfunction. In the
model proposed, death of the organism is associated with a threshold
beyond which this organism dies, which can be interpreted in the manner
of hyperfunction theory as excessive functions of gene products for
positive leaders of aging direction. However, some significant differences
from hyperfunction theory as proposed in \cite{blagosklonny2006aging,blagosklonny2009growth}
should be noted. Namely, we do not have any experimental data to determine
the role of aging direction in developmental processes so this connection
remains speculative. Moreover, according to the theoretical model,
changes along stable developmental directions will tend to diminish
with time and will not contribute to changes in gene expression, so
at least some fraction of genes involved in development will not contribute
to aging. 
\item \textbf{damage/error accumulation} \cite{Partridge2002,Sinclair2009,Gladyshev2012,Gladyshev2013}.
The stochastic component of the expressome $\delta x(t)$ can be interpreted
as a result of accumulation of errors, e.g. regulatory errors. Though
it becomes small compared to the deterministic component at late ages,
its effects are strongly pronounced at all ages, as the mean square
deviation $\sqrt{E(\delta x^{2}(t))}$ first grows with age at the
same exponential Gompertz rate as the mortality $M(t)$ and later
stabilizes at some constant level, contributing at late ages to the
very process of organism passing beyond toxicity threshold and dying.
Our analysis shows, that the stochastic forces shaping the expressome
evolution early in life explain a wide distribution of the ages at
death even in a genetically homogenous population. It should be noted
that damage accumulation does not exclude the scenario of quasi-programmed
aging, as damage is not limited to byproducts, errors and other molecular
forms, but encompasses deleterious changes at all levels, or the deleteriome
\cite{Gladyshev2012,Gladyshev2013,gladyshev2016aging}. 
\end{itemize}
The relationship just established between the critical dynamics of
the gene regulatory network state vector and the Gompertzian mortality
characteristic of most species allows one to consider the projection
$z=(b^{T}\cdot\delta x)$ as a candidate biomarker of aging. Due to
the low value of the eigenvalue $\epsilon$ the direction $b$ dominates
the response to any generic stress. This, in fact, was already demonstrated
in \cite{10.1371/journal.pone.0086051}, where transcriptional signatures
of responses to very different stresses were found to share a great
number of gene expression changes. 

The levels of gene expression or metabolites corresponding to the
most significant components of the aging direction $b$ are strongly
associated with age and age-related diseases. Therefore aging as described
here, in terms of GRN instability, leads to the impairment of normal
physiological functions, has characteristic biomarkers and phenotypes
including signs of major diseases of aging, and produces exponentially
increasing morbidity. Accordingly, the process of aging itself falls
under the definition of a disease recently used by AMA for a common
condition such as obesity \cite{american2013resolution}. We must
note here that therapeutic or experimental interventions aimed to
counter-balance the age-related changes in gene or metabolite expression
may not be very effective ways to extend lifespan of the species.
For example, even though aging in \emph{D. melanogaster} is associated
with increased internal and external bacterial load, as well as with
increased expression of antibacterial peptides, neither reducing the
bacterial population with antibiotics nor reducing the humoral antibacterial
response was able to extend lifespan \cite{ren2007increased}. The
transcription factor $hairy$ is overexpressed with age, and yet its
inhibition does not result in lifespan increase \cite{zhou2008mechanisms}.
This reinforces the conclusion that the markers of aging are, in general,
not the same as regulators of aging.

Even though most of the analysis in this manuscript is performed for
\emph{D. melanogaster}, our conclusions are generic and should be
applicable to other species. Depending on the network and environmental
parameters, realistic GRNs would not necessarily support a derivation
of the Gompertz law. In fact, aging is extremely diverse across the
tree of life \cite{jones2014diversity}. We believe that one of the
most intriguing form of a non-Gompertzian mortality law in the model
may arise if the effective potential $V(z)$ has a local minimum with
small but positive curvature, as in Figure \ref{fig:SchematicVandPCA}A,
$\alpha<0$. The higher order nonlinearities, such as the cubic terms
in the effective potential, cannot be neglected anymore and in this
case the genetic network turns out to be metastable. If the minimum
of the potential $V(z)$ is separated from the region of large $z$
by a sufficiently high activation barrier\textcolor{black}{{} (see Figure
\ref{fig:SchematicVandPCA}A and Supplementary Information, Appendix
\ref{subsec:gene-network-kinetics})}, the mortality rate, determined
by the probability of activation, is exponentially small and age-independent.
This situation could be a physical picture behind the hypothesis that
some animal species exhibit negligible senescence \cite{finch1994longevity}
This feature of our model may also explain the apparent lack of age-dependent
changes in physiological parameters as well as mortality rates observed
in the naked mole rat and some other species over long time periods
\cite{buffenstein2005naked}. This argument may also be supported
by the analysis in \cite{kim2011genome,loram2012age-related}, where
the number of the genes differentially expressed with age was compared
between extremely long-lived species and mice and humans. These works
show that the number of differentially expressed genes in species
with negligible senescence is much lower than in mammals with Gompertzian
aging.
\begin{acknowledgments}
The authors are grateful to Profs. A. Moskalev and V. Fontana, and
Drs. A. Avanesov and M. Konovalenko for valuable discussions, Dr.
U. Fischer, S. Filonov, M. N. Kholin, and Dr. S.W. Barger for enlightening
comments, discussions and substantial help in preparation of the manuscript.
We would also like to thank Prof. J. Carey and Prof. S. Pletcher for
access to their experimental data.
\end{acknowledgments}

\bibliographystyle{plain}
\bibliography{Qrefs,DP}

\appendix

\section*{Supplementary Information}

\section{Stochastic dynamics of gene expression and metabolite levels \label{subsec:Slow-stochastic-dynamics-1}}

In the following, we explain the physical basis for Eq. (\ref{eq:1DoF})
and provide its derivation. We also explicitly find the expressions
for the mortality rate, the survival probability, the lifespan distribution
function and the average lifespan of species with expressome subject
to Eq. (\ref{eq:1DoF}). 

\subsection{Getting a grasp on expressome dynamics. Model reduction and physical
considerations}

\begin{figure}
\includegraphics[width=0.5\textwidth]{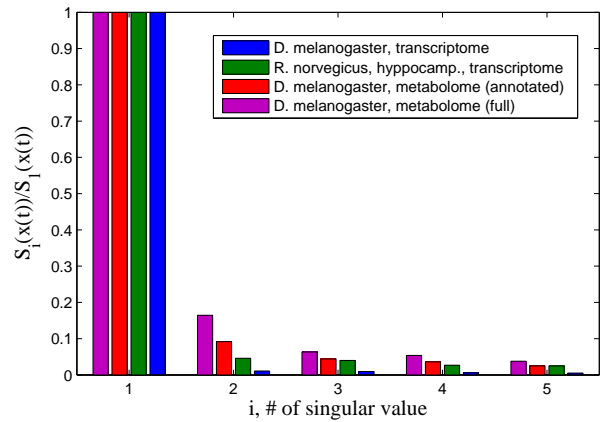}\caption{Five largest singular values $S_{i}(x(t))$ of the expressome $x(t)$
divided by the largest one, $S_{1}(t)$; transcriptome of \emph{D.
melanogaster} \cite{pletcher2002genome} (blue), transcriptome of
\emph{R. norvegicus,} hippocampus \cite{kadish2009hippocampal} (green),
metabolome of \emph{D. melanogaster, }targeted metabolites only \cite{avanesov2014age}
(red), metabolome of \emph{D. melanogaster,} all metabolites \cite{avanesov2014age}
(purple). One of the orthogonal components of the expressome $x(t)$
clearly dominates over the others in all considered cases.\label{fig:Five-largest-SVD}}
\end{figure}

The analysis represented in this paper is mainly based on the observation
that time series datasets describing changes of expressome profiles
with age are often quite susceptible to Model Reduction techniques
\cite{Antoulas2009}. Proper orthogonal decomposition of dynamic (time-series)
datasets for age-associated gene expression levels (and also metabolite
levels or other \textquotedblleft omics\textquotedblright{} datasets)
\cite{pletcher2002genome,kadish2009hippocampal,avanesov2014age} shows
that the long-time behavior of any expressome $x$ is largely determined
by a single component (see Fig. \ref{fig:Five-largest-SVD}). The
number of different principal components $PC_{n}$ representing the
signal $x(t_{1}),\ldots,x(t_{m})$ cannot not exceed the number of
time snapshot points in corresponding time series. Even though $m$
is typically limited and is very small compared to the number of transcripts
or metabolites, , the clear dominance of a single component in the
signal is a non-trivial fact. As we argue here, it carries important
information about biological gene regulatory and metabolic networks;
namely, it shows that such networks are pre-critical in a sense that
will be explained below.

Consider a meta-stable gene regulatory network (GRN) with the instantaneous
network state represented by a vector $x$, whose components $x_{i}$
are given by expression levels of different genes, proteins or metabolites
participating in the GRN. Dynamics of the vector $x$ consists of
mRNA levels, concentrations of proteins and metabolites, involved
in regulatory pathways and are influenced by external as well as internal
stress factors acting on the cell. This dynamics is governed by differential
matrix equations of systems biology

\begin{equation}
g\left(x,\frac{dx}{dt},\frac{d^{2}x}{dt^{2}},\ldots\right)=F,\label{eq:generic-eq0}
\end{equation}
where $g$ is a vector function of the state vector $x$, encoding
interactions between different components of the expressome, and the
vector $F$ describes the action of (mostly stochastic) external or
internal stress factors affecting components of $x$. For the sake
of simplicity, below we will refer to gene expressions only, if not
stated otherwise. 

Generally speaking, the vector function $g$ may depend on arbitrarily
high time derivatives $\frac{dx}{dt}$, $\frac{d^{2}x}{dt^{2}},\ldots,\frac{d^{n}x}{dt^{n}},\ldots$
of the expressome state vector. Hereinafter the focus of the study
is on the dynamics of $x$ at time scales comparable to the mortality
rate doubling time $t_{{\rm MRDT}}$, which is of the order of tens
of days for nematodes and drosophilae and hundreds of days for mice.
This time scale is to be compared to a typical time scale of protein
translation and regulation, or a lifetime of mRNA. In order to describe
this \emph{slow }dynamics of the expressome state vector $x$, one
can neglect all time derivatives of $x$ higher the first derivative
$\frac{dx}{dt}$. The matrix Eq. (\ref{eq:generic-eq0}) thus reduces
to
\begin{equation}
g\left(x,\frac{dx}{dt}\right)=F.\label{eq:generic-eq}
\end{equation}

\subsection{Frequency domain properties of the vector $F$ of stress factors}

As it was explained in the main text, the vector $F$ of environmental
and internal stress factors affecting gene expression levels $x$
has both a slowly changing component, $F_{0}(t)$, and a stochastic
component, $\delta F(t)$, rapidly changing in time. Below we assume
for simplicity that $F_{0}(t)=F_{0}$, i.e., the slowly changing component
is, in fact, constant. 

Since dynamics of the vector $\delta F$ can be considered essentially
random at time scales of interest, dynamics of $x$ is driven by correlation
properties of $\delta F$. The total number of environmental and internal
stress factors affecting gene expression levels $x$ is extremely
large, and the vector $\delta F(t)$ represents a superposition of
these factors. According to the central limit theorem \cite{Schervish2011}
it can thus be considered a Gaussian stochastic process, so that 
\[
E(\delta F(t)\delta F(t'))=B(t-t'),
\]
where $B(t-t')$ is a function of the time difference $|t-t'|$ only,
while $E(\ldots)$ stands for the expected value of its argument.
The statistical average is understood as average over ensemble of
species/cells of a given organism. In the frequency domain, one has
correspondingly
\[
E(\delta F(f)\delta F(-f))=
\]
\[
=\int df\,e^{-2\pi if(t-t')}E(\delta F(t)\delta F(t'))=B(f),
\]
where $f$ is frequency of a given mode in the Fourier expansion of
the function $B(t-t').$ 

The function $B(f)$ has multiple singularities in the complex plane
of $f$. These singularities encode characteristic time scales of
pathways' dynamics. However, the function $B(f)$ decays quickly with
$f$ as $f\to\infty$. This guarantees that the amount of stress affecting
the gene expression levels $x$ does not change very rapidly. On the
other hand, the behavior of $B(f)$ is smooth at very small frequencies,
i.e., $B(f)\to B$ as $f\to0$, so that the amplitude of rare fluctuations
of $\delta F$ remains limited. As we would like to understand the
dynamics of $x$ at time scales comparable to $t_{{\rm MRDT}}$, we
consider the case $B(f)\sim B$ in what follows. In this, the stochastic
process $\delta F(t)$ in the right-hand side of Eq. (\ref{eq:generic-eq})
has correlation properties of white noise:

\begin{equation}
E(\delta F(t)\delta F(t^{\prime}))=B\delta(t-t^{\prime}),\label{eq: Noise corellation properties}
\end{equation}
where $\delta(t-t')$ is the Dirac delta-function.

\subsection{The vicinity of a stationary point (homeostasis)}

When the fluctuations of the genotoxic forces are relatively weak,
$|\delta F|\ll\left|F_{0}\right|$, the fluctuations of the expressome
state vector are also small: $x=x_{0}+\delta x$, $\delta x\ll x_{0}$.
Here $x_{0}$ is the stationary point given by the solution of the
equation 
\[
g(x_{0})=F_{0}.
\]
In the vicinity of this point Eq. (\ref{eq:generic-eq}) reduces to
\begin{equation}
D\delta\dot{x}+K\delta x+\Gamma^{(3)}\delta x\delta x+\ldots=\delta F,\label{eq: Main equation}
\end{equation}
where the matrices $D$ and $K$ describe the relaxation effects and
the ``interaction'' between genes in the GRN. The matrix $\Gamma^{(3)}$
encodes the leading non-linear terms in the expansion of the vector
function $g(x)$ in powers of small $\delta x$, with all other terms
corresponding to the higher order non-linearities being omitted from
the expansion. Generally, such non-linear terms are not small (in
particular, they are not small at early times, close to the initial
state of the expressome). However, a common feature of the dissipative
network described by Eq. (\ref{eq: Main equation}) is the decay of
(most) components of $\delta x$ with time, which makes non-linearities
in Eq. (\ref{eq: Main equation}) less relevant at later stages of
the time dynamics. Whether this feature holds for the GRN under consideration
is a non-trivial question, which we shall now address.

\subsection{Saddle-node bifurcation and critical slowing-down}

Domination of a single principle component in the expressome vector
$x(t)$ is \emph{not} a generic prediction of the model (\ref{eq: Main equation}).
We would like to argue that such a dominance implies that GRNs under
consideration are operating at the critical point. This critical point
is a bifurcation, separating regimes of stable and unstable run-away
behavior of the expressome $\delta x$.

Transition from stability to instability in networks with network
graphs not possessing specific symmetries is typically associated
with existence of co-dimension 1 bifurcations \cite{Fiedler2002,Seydel2009}.
Such transition is characterized by the loss of stability along a
single orthogonal component of the network state vector. Contributions
from all other orthogonal components remain stable. This situation
is known in the literature as saddle-node bifurcation \cite{Fiedler2002}
and is realized when the lowest (real) eigenvalue of the matrix $K$
reaches zero and then becomes negative. 

Let us denote the smallest (vanishing, but positive) eigenvalue of
the matrix $K$ as $\epsilon$ and its corresponding left and right
eigenvectors as $a$ and $b$. In order to understand dynamics of
the expressome $\delta x$ near the GRN bifurcation, it is convenient
to analyze the autocorrelation function $E(\delta x(t)\delta x(t'))$.
As shown in \cite{podolsky2013critical,podolsky2013random}, the latter
is given by the following expression (in the frequency domain)

\[
E(\delta x(t)\delta x^{T}(t'))=
\]
\begin{equation}
=\int df\cdot e^{-2\pi if(t-t')}A^{-1}(f)E(\delta F(-f)\delta F^{T}(f))A^{*,-1}(f),\label{eq:CorrFunctionDef}
\end{equation}
where $A(f)=i2\pi Df+K$. Its behavior is determined by the poles
of the integrand in the complex plane of frequency $f$. In turn,
the latter are given by the solution of the equation 
\[
\det(A(f))=0,
\]
i.e., by the solution of the eigenvalue problem for the matrix $\pm i(2\pi)^{-1}D^{-1}K$.
In particular, at very large time separations $|t-t'|$ one finds
that 
\begin{equation}
E(\delta x(t)\delta x^{T}(t'))\approx\frac{(a^{T}\cdot Ba)b\cdot b^{T}}{\epsilon(a^{T}\cdot Db)^{3}}\exp\left(-\frac{\epsilon|t-t'|}{(a^{T}\cdot Db)}\right).\label{eq:CorrFunctionStable}
\end{equation}
Several important observations can be made at this point: 1) as the
parameter $\epsilon$ approaches $0$, amplitude of the autocorrelation
function (\ref{eq:CorrFunctionStable}) grows as $\epsilon^{-1}$,
i.e. fluctuations of the expressome vector are strongly amplified
both with age and among the population (the statistical ensemble);
2) the autocorrelation time scale $\tau=(a^{T}\cdot Db)/\epsilon$
also grows strongly at $\epsilon\to0$, implying that stochastic dynamics
of the expressome becomes very slow near the point of bifurcation,
a phenomenon known as critical slowing-down (see for example \cite{Suzuki2009});
3) fluctuations of the expressome vector $\delta x$ mostly develop
along the direction $b$ of the right eigenvector of $K$ corresponding
to the vanishing eigenvalue $\epsilon$, which in its turn guarantees
that a single principal component of the expressome vector $\delta x$
dominates; 4) the corresponding left eigenvector $a$ determines directions
of high sensitivity of the expressome to genotoxic stress factors
$\delta F$. Other remaining eigenvalues of $-i(2\pi)^{-1}D^{-1}K$
have strictly positive real parts and thus correspond to contributions
to (\ref{eq:CorrFunctionStable}), rapidly decaying with time. This
guarantees that only one orthogonal component of $\delta x$ dominates,
in accordance with experimental observations.

It should be noted that, when higher order derivatives in Eq. (\ref{eq:generic-eq0})
are taken into account, this argument, albeit becoming more involved,
remains valid: one has to construct the perturbation theory in powers
of small parameter $\epsilon$, which still determines the smallest
eigenvalue of the matrix $A(f)=K+i2\pi fD+(2\pi f)^{2}M+\ldots.$
Similarly to the case considered here, all higher eigenvalues of $A(f)$
possess positive real parts. 

When $\epsilon$ crosses $0$ and becomes negative, the autocorrelation
function (\ref{eq:CorrFunctionStable}) develops unstable behavior,
again directed along the corresponding right eigenvector $b$ of $K$.
In particular, one finds for the dispersion of the expressome at large
$t$
\begin{equation}
E(\delta x(t)\delta x^{T}(t))\approx E(z^{2}(0))b\cdot b^{T}\exp\left(\frac{2\epsilon t}{(a^{T}\cdot Db)}\right),\label{eq:StochasticGrowth}
\end{equation}
where $E(z^{2}(0))$ is the dispersion of the projection $z=(\delta x^{T}\cdot b)$
of the expressome on the vector $b$, taken at the initial moment
of time. The displacements of the expressome $\delta x$ along the
other orthogonal components correspond to the dynamics of the gene
network along the eigenvectors, characterizing the modes with finite
and positive eigenvalues, and hence remain stable in time.

Thus, the quantity of interest in the critical regime $\epsilon\to0$
is the projection of the expressome $\delta x$ on the right eigenvector
$b$, $z=(\delta x^{T}\cdot b),$ which satisfies the Langevin equation
\begin{equation}
\frac{dz}{dt}=v\left(z\right)+\delta F',\label{eq:LangevinDerived}
\end{equation}
where $v\left(z\right)$ is the ``velocity'', characterizing the
motion along the $z$ coordinate. Next to the origin the velocity
$v(z)\approx\alpha z$, where $\alpha=-\epsilon/(a^{T}\cdot Db)$.
Here $\delta F'=F/(a^{T}\cdot Db)$, is the stochastic force, such
that $E(\delta F'(t)\delta F'(t'))=\Delta\delta(t-t')$, and $\Delta=(a^{T}\cdot Ba)/(a^{T}\cdot Db)^{3}$
is a property of the fluctuations. Since the vector $b$ is defined
as eigenevector and subsequently is defined up to a factor, we can
always select its direction in such a way as to make the collective
coordinate $z$ increase with the age. It means that we can always
limit our attention to the region $z>0$ only. 

\section{Estimating mortality rates from the stochastic dynamics of an expressome
\label{subsec:gene-network-kinetics}}

So far we have only considered stochastic dynamics of the expressome
state vector $\delta x$ of an organism. To proceed with population
properties, such as mortality, we need to construct a statistical
description for populations, ensembles of organisms. Let us consider
a large population of animals with the number of individuals $N(t=0)=N_{0}\gg1$,
born at the same moment of time, $t=0$. The development and aging
of the animals in the population depends on the dynamics of the expression
levels $x$ and can be naturally described in terms of the probability
density $dN=N_{0}P\left(x,t\right)dx$, where $P(x,t)$ is the fraction
of individuals within the population, characterized by the gene expression
levels from the interval $(x,x+dx)$, and estimated at the time $t$.
Another quantity of interest is the survival probability 
\[
S(t)=\int P\left(x,t\right)dx,
\]
defined as the fraction of the animals, $N(t)/N_{0}$, surviving by
the age $t$, relative to the initial size of the population $N_{0}$.
The mortality rate can be found using the standard definition $M(t)=-(dS/dt)/S.$
In this Section we shall explain how to estimate the probability density
$P(x,t)$, the survival probability $S(t)$, and the mortality rate
$M(t)$ from the stochastic dynamics of the expressome state vector
$\delta x.$ 

\subsection{Estimating the probability density $P(z,t)$}

\subsubsection{Fokker-Planck equation }

Following the arguments above we describe the process of aging as
the slow changes in gene expression levels $\delta x$ over time governed
by the Langevin equation (\ref{eq:LangevinDerived}). The probability
$P(z,t)$ is the solution of the associated Fokker-Planck equation
\cite{Zwanzig2001} 
\begin{equation}
\frac{\partial P(z,t)}{\partial t}+\frac{\partial J\left(z,t\right)}{\partial z}=0,\label{eq:FokkerPlanck}
\end{equation}
where
\[
J\left(z,t\right)=v(z)P(z,t)-\frac{1}{2}\Delta\frac{\partial P(z,t)}{\partial z}
\]
is the probability flux along the direction $z$. We seek the distribution
function $P(z,t)$ as the solution, corresponding to the initial condition
$P(z,t=0)=P_{0}(z)$, and boundary conditions at $z=0$ and large
$z$. To exclude the $z<0$ region we place a reflecting wall at $z=0$,
which corresponds to
\begin{equation}
\left(\frac{\partial P(z,t)}{\partial z}\right)_{z=0}=0.\label{eq: Reflecting wall at z=00003D0}
\end{equation}
The suggested form of the boundary condition means that the $z>0$
and $z<0$ regions are totally isolated from each other. 

We explore a Gaussian form of the initial condition in the form
\begin{equation}
P_{0}(z)=G\left(z\right)+G\left(-z\right),\label{eq:FPGaussianInitialCondition}
\end{equation}
\[
G\left(z\right)=\frac{1}{\sqrt{2\pi}\sigma_{0}}e^{-\frac{(z-z_{0})^{2}}{2\sigma_{0}^{2}}},
\]
characterized by the initial displacement $z_{0}$, and the distribution
width, $\sigma_{0}$. The probability distribution (\ref{eq:FPGaussianInitialCondition})
is an even functions and hence satisfies the boundary condition for
(\ref{eq: Reflecting wall at z=00003D0}) automatically. Therefore,
the normalization condition takes the form 
\begin{equation}
\intop_{0}^{\infty}dzP_{0}(z)=1.\label{eq: Normalization}
\end{equation}
In fact, in many practical situations Eq. (\ref{eq:FokkerPlanck})
can be solved without the the boundary condition (\ref{eq: Reflecting wall at z=00003D0})
using a simpler form of the initial distribution $P_{0}(z)=G\left(z\right)$.
In this case the fraction of the animals

\[
P_{-}=\frac{1}{2}\left[1-{\rm erf}\left(\frac{z_{0}}{\sqrt{2}\sigma_{1}}\right)\right],\quad\sigma_{1}=\sqrt{\sigma_{0}^{2}+\frac{\triangle}{2\alpha}},
\]
will be, of course, lost to the unphysical region $z<0$. Therefore,
the results of such a simplified calculation could only be trusted
once the ``leak'' is sufficiently small, $P_{-}\ll1$, or, equivalently,
if $z_{0}\gg\sigma_{1}$. We distinguish between the two quantitatively
different types of the initial conditions: the ``deterministic'',
\begin{equation}
z_{0}\gg\sigma_{1},\label{eq: Motion only to the right}
\end{equation}
and the ``diffusive'', $z_{0}\ll\sigma_{1},$forms respectively. 

For sufficiently small values of $z$ the ``velocity'' term in the
Langevin equation (\ref{eq:LangevinDerived}) can be approximated
by its linear expansion

\begin{equation}
v\left(z\right)=\alpha z.\label{eq: small nonlinearity}
\end{equation}
 For larger $z$ the nonlinearities set in and we employ a more general
form 
\begin{equation}
v\left(z\right)=\alpha z\phi\left(z\right),\label{eq: General form of v versus z}
\end{equation}
where $\phi\left(0\right)=1$ is chosen to match the asymptotic expression
(\ref{eq: small nonlinearity}) for small $z$. More specifically
we expect that the function $\phi\left(z\right)$ increases at $z\rightarrow\infty$
fast enough, so that that $\ln(z)/\phi\left(z\right)\rightarrow0$
(see the discussion below). 

Biological states characterized by extremely large values of $z$
correspond to highly distorted expression profiles and therefore we
assume, that the state $z=\infty$ is certainly associated with the
death of the individual. Therefore, a physically relevant choice of
the boundary condition at large $z$ is
\begin{equation}
P(z\to\infty,t)=0.\label{eq:BoundaryConditionNlin}
\end{equation}
The Fokker-Planck equation (\ref{eq:FokkerPlanck}) with the boundary
condition in the form of Eq. (\ref{eq:BoundaryConditionNlin}) cannot
be solved exactly. Fortunately, this is not really needed: a large
amount of useful information about the stochastic dynamics of the
expressome in relation to aging can be extracted from the asymptotic
behavior of the probability density $P(z,t)$ in a few physically
interesting regimes. The following analysis of the solutions of Eq.
(\ref{eq:FokkerPlanck}) is not, of course, new and we loosely follow
the results presented in \cite{Suzuki1977,Suzuki2009} in the remaining
of manuscript. 

\subsubsection{A toy model: absorbing wall at a large $z=Z$ }

Let us first collect a few well known results about the solutions
of Eq. (\ref{eq:FokkerPlanck}) for small $z$, where the drift velocity
$v(z)$ can still be used in its linear form (\ref{eq: small nonlinearity}).
If, for simplicity of the argument, the initial distribution is very
narrow, $\sigma_{0}\rightarrow0$ and $z_{0}=0$ ($G(z)\approx\delta(z)$),
then the solution satisfying the boundary condition (\ref{eq:BoundaryConditionNlin})
is
\begin{equation}
P\left(z,t\right)=2G\left(z,0,t\right)=2\sqrt{\frac{\alpha}{\pi X\left(t\right)\varDelta}}\exp\left[-\frac{\alpha z^{2}}{X\left(t\right)\varDelta}\right],\label{eq: Solution of linear with the absence of the absorbing wall}
\end{equation}
where 

\[
G\left(z,z^{\prime},t\right)=\sqrt{\frac{\alpha}{\pi X\left(t\right)\varDelta}}\exp\left[-\frac{\alpha\left(z-z^{\prime}e^{\alpha t}\right)^{2}}{X\left(t\right)\varDelta}\right]
\]
is the Green function of Eq. (\ref{eq:FokkerPlanck}), and $X\left(t\right)=\exp\left(2\alpha t\right)-1$.
Therefore, at sufficiently early times, $t\alt\alpha^{-1}$, the effect
of the drift term is negligible and the behavior of $z$ is described
by the diffusion equation with $v(z)\approx0$. This means that at
least in the beginning the solution is dominated by the effect of
the random force $f(t)$ over the ``deterministic'' force $-\alpha z$
\cite{Suzuki1977,Suzuki2009}. 

Over time the deterministic force is taking over. It is instructive
to observe the relevant regimes of the solution using a toy model
first. To emulate the effect of the nonlinearity let us choose the
form of the boundary condition at large $z$ in the form of a totally
absorbing wall,
\begin{equation}
P(z=\pm Z,t)=0,\label{eq:BoundaryConditionLin}
\end{equation}
suggesting that the an organism dies whenever its $z$ reaches a certain
threshold value $Z$. The solution (\ref{eq: Solution of linear with the absence of the absorbing wall})
remains valid for sufficiently small ages, 
\begin{equation}
t\alt\frac{1}{\alpha}\ln\left(Z\sqrt{\frac{\alpha}{\varDelta}}\right)\equiv t_{ls},\label{eq: Age of achieving the absorbing wall}
\end{equation}
before the appreciable fraction of the animals reaches the $z\sim Z$,
and hence $t=t_{ls}$ is of order of the lifespan in the population.
Often the life expectancy $t_{ls}$ exceeds the GRN instability time
scale, $t_{\alpha}\sim1/\alpha$, or 

\begin{equation}
\frac{\alpha Z^{2}}{\varDelta}\agt1.\label{eq: Main large parameter}
\end{equation}
It is shown below, that $\nicefrac{\alpha Z^{2}}{\varDelta}$ is the
basic large parameter of our model. Whenever the ratio of the time
scales is sufficiently large, the effects of the non-linearities are
``weak'' and a complete analytical treatment of the problem becomes
possible and yields a series of universal results. 

Technically, to meet the boundary condition (\ref{eq:BoundaryConditionLin})
requirement, we add a loss term at the position of the absorbing wall,$-\eta\left(t\right)\left[\delta\left(z-Z\right)+\delta\left(z+Z\right)\right]$,
where $\eta\left(t\right)$ is the Lagrange multiplier, an auxiliary
function introduced to enforce the constraint (\ref{eq:BoundaryConditionLin}).
The analysis of the modified equation is straightforward and yields
\[
P\left(z,t\right)=\varPhi\left(z,t\right)-
\]
\begin{equation}
-\intop_{0}^{t}\left[G\left(z,Z,t-t^{\prime}\right)+G\left(z,-Z,t-t^{\prime}\right)\right]\eta\left(t^{\prime}\right)dt^{\prime},\label{eq: The solution for the linear with absorbing walls}
\end{equation}
where
\[
\varPhi\left(z,t\right)=\intop_{-\infty}^{+\infty}G\left(z,z^{\prime},t\right)P_{0}\left(z^{\prime}\right)dz^{\prime}.
\]
If the non-linearity is weak or the wall is far from current state
of the system, i.e. whenever (\ref{eq: Main large parameter}) holds,
then $G\left(Z,-Z,t-t^{\prime}\right)\ll G\left(Z,Z,t-t^{\prime}\right)$,
the function $\eta\left(t^{\prime}\right)$ varies very slowly with
age compared to $G\left(Z,Z,t-t^{\prime}\right)$, and therefore 
\begin{equation}
\eta\left(t\right)\approx\alpha Z\varPhi\left(Z,t\right).\label{eq: Function g}
\end{equation}
For the case of Gaussian initial distribution (\ref{eq:FPGaussianInitialCondition})
we have 
\begin{equation}
\varPhi(z,t)=\frac{1}{\sqrt{2\pi}\sigma\left(t\right)}\left[e^{-\frac{(z-z\left(t\right))^{2}}{2\sigma^{2}\left(t\right)}}+e^{-\frac{(z+z\left(t\right))^{2}}{2\sigma^{2}\left(t\right)}}\right],\label{eq:ProbEarlyTime}
\end{equation}
where $z\left(t\right)=z_{0}\exp(\alpha t)$, and the effective distribution
width increases exponentially with age
\begin{equation}
\sigma^{2}\left(t\right)=\sigma_{0}^{2}+\sigma^{2}X\left(t\right),\label{eq:Disp-1}
\end{equation}
where the combined quantity
\[
\sigma=\sqrt{\sigma_{0}^{2}+\frac{\Delta}{\alpha}}
\]
results from the initial probability density width broadening by diffusion
at small $t$.

\subsubsection{Weak non-linearity limit, general case}

Let us turn our attention back to the general form of the non-linear
drift velocity (\ref{eq: General form of v versus z}).At sufficiently
small $z$ function $v(z)$ can be approximated by its linear expansion
$v(z)\approx\alpha z$ and therefore Eq. (\ref{eq:FokkerPlanck})
can be studied in its simplified form
\begin{equation}
\frac{\partial P(z,t)}{\partial t}+\frac{\partial}{\partial z}\left[\alpha zP(z,t)-\frac{1}{2}\Delta\frac{\partial P(z,t)}{\partial z}\right]=0.\label{eq: Fokker Planck in the linear region}
\end{equation}
We introduce a characteristic length scale $z\sim Z$, where the linear
approximation to $v(z)$ fails. The age when $z\sim Z$ will correspond
roughly to the lifespan of the organism, since, according to our assumptions,
$v(z)$ grows fast enough at large $z$ and the point $z=\infty$
is reached in a finite time.

This linearized equation can be solved exactly with initial distribution
(\ref{eq:FPGaussianInitialCondition}) and boundary condition (\ref{eq:BoundaryConditionNlin}):

\begin{equation}
P(z,t)=\frac{1}{\sqrt{2\pi}\sigma}\exp\left\{ -\alpha t-\frac{\left(ze^{-\alpha t}-z_{0}\right)^{2}}{2\sigma^{2}}\right\} \label{eq: Solution for lin FP with diffusion}
\end{equation}
which would be a rightful approximation for solution of Eq. (\ref{eq:FokkerPlanck})
in time region $t\ll t_{ls}$.

In the same time, at $t\gg\alpha^{-1}$ , irrespectively of the specific
form of the nonlinearity, one can neglect the diffusion and set $\varDelta=0$.
This means that the dynamics of the expressome is dominated by the
drift term and is, therefore, deterministic. Accordingly, Eq. (\ref{eq:FokkerPlanck})
can be transformed into 
\begin{equation}
\frac{\partial P(z,t)}{\partial t}+\frac{\partial}{\partial z}\left[v\left(z\right)P(z,t)\right]=0.\label{eq: Drift equation}
\end{equation}
One of the consequences of Eq.(\ref{eq: Drift equation}) is the fact
that if a specimen is observed with an expressome state, corresponding
to $z\gg\sqrt{\varDelta/\alpha}$, then the remaining life expectancy,
or the time-to-death, for the animal is 
\begin{equation}
\tau\left(z\right)=\intop_{z}^{\infty}\frac{dz^{\prime}}{v\left(z^{\prime}\right)},\label{eq: time-to-death}
\end{equation}
and is finite. The integral on the r.h.s. is logarithmically divergent
for the chosen type of non-linearity \ref{eq: General form of v versus z},
which means that 
\[
\tau(z)=\frac{1}{\alpha}\ln\left(\frac{1}{z}\right)+\widetilde{\tau}\left(z\right),
\]
and 
\[
\underset{z\rightarrow0}{\lim}\widetilde{\tau}\left(z\right)={\rm const}
\]
is finite. 

The solution of Eq.(\ref{eq: Drift equation}) takes form $P(z,t)=P(z,\xi)=C(\xi)/\nu(z),$
where $\xi=t+\tau(z)$ and $C(\xi)$ is a function dependent on initial
distribution and boundary conditions. One way to find it is to match
the solutions of Eqs. (\ref{eq: Fokker Planck in the linear region})
and (\ref{eq: Drift equation}) at some intermediate age $t_{1}$,
lying in time region, where both equations can be used. Indeed, since
normally $\alpha^{-1}\ll t_{ls}$ , we can use any $t_{1}$ such that 

\begin{equation}
\alpha^{-1}\ll t_{1}\ll t_{ls}\label{eq:CommonAgeRange}
\end{equation}
as the matching point. The age range corresponds to $\sqrt{\Delta/\alpha}\alt z_{1}\alt Z$
interval. As the result, we obtain a closed form solution, applicable
for deterministic initial condition (\ref{eq: Motion only to the right})
and boundary condition (\ref{eq:BoundaryConditionNlin}) for all ages
$t\gtrsim1/\alpha$
\begin{equation}
P(z,t)=\frac{\alpha k\left(z\right)}{\sqrt{2\pi}\sigma v\left(z\right)}\exp\left\{ -\alpha t-\frac{\left[k\left(z\right)e^{-\alpha t}-z_{0}\right]^{2}}{2\sigma^{2}}\right\} .\label{eq: P universal weak non-linearity}
\end{equation}
Here

\begin{equation}
\widetilde{Z}=\underset{z\rightarrow0}{\lim}e^{\alpha\widetilde{\tau}\left(z\right)}\sim Z,\label{eq: Definition of L}
\end{equation}
and 
\begin{equation}
k\left(z\right)=\widetilde{Z}\exp\left[-\alpha\tau\left(z\right)\right]\label{eq: Definition of k}
\end{equation}
is auxiliary function with known asymptotic forms for the advanced
($\tau\left(z\right)\rightarrow0$)
\begin{equation}
k\left(z\right)\rightarrow\widetilde{Z}\quad at\;z\rightarrow\infty,\label{eq: k limit at infty-1}
\end{equation}
and the early ($\tau\left(z\right)\rightarrow\infty$) 
\begin{equation}
k\left(z\right)\rightarrow0\quad at\;z\rightarrow0\label{eq: k limit at zero-1}
\end{equation}
ages.

Let us turn to a more specific case of an arbitrary polynomial non-linearity
\begin{equation}
\phi\left(z\right)=1+\left(\frac{z}{Z}\right)^{\beta}\label{eq: More general case than linear}
\end{equation}
with $\beta>0$. The length scale here characterizes the non-linearity
strength and hence large values of $Z$ correspond to the weak non-linearity
limit (\ref{eq: small nonlinearity}). In accordance with the earlier
definitions, it can be shown that $\tilde{Z}$ equals $Z$, and the
remaining lifespan is 
\[
\tau\left(z\right)=\frac{1}{\alpha}\ln\left\{ \frac{\left[Z^{\beta}+z^{\beta}\right]^{1/\beta}}{z}\right\} ,
\]
and the probability distribution function takes the form of Eq. (\ref{eq: P universal weak non-linearity})
with with $k\left(z\right)=z/\left[1+(z/Z)^{\beta}\right]^{1/\beta}$. 

The absorbing wall at $z=Z$ situation, considered in the previous
section, formally corresponds to $\beta=\infty$. Interestingly, the
solution is still valid, $k(z)=z$ everywhere, if not immediately
close to the wall, i.e. Eq. (\ref{eq: P universal weak non-linearity})
matches the exact expression (\ref{eq: The solution for the linear with absorbing walls})
for all $z\alt Z$. Hence we can conclude that the derived form of
the probability density is universal, holds for an arbitrary form
of the non-linearity, allowing for a finite lifespan. A specific form
of the non-linearity is not important, all the information about the
non-linear terms can be compressed into the length scale $Z$, being
the single important parameter defining the lifespan and the form
of the distribution function in the population. 

Let us finish the section by computing the fraction of the animals,
surviving by the age $t$. According to the definition, 
\[
S\left(t\right)=\intop_{0}^{+\infty}dzP\left(z,t\right)=
\]
\[
=\frac{1}{\sqrt{2\pi}\sigma}\intop_{0}^{Z}dk\exp\left\{ -\alpha t-\frac{\left[k\left(z\right)e^{-\alpha t}-z_{0}\right]^{2}}{2\sigma^{2}}\right\} =
\]
\[
=\frac{1}{\sqrt{\pi}}\intop_{y_{1}}^{y_{2}}dye^{-y^{2}},
\]
where, for $\alpha t\gg1$, 
\[
y_{2}=\frac{Ze^{-\alpha t}-z_{0}}{\sqrt{2}\sigma},\;y_{1}=\frac{-z_{0}}{\sqrt{2}\sigma}.
\]
That would lead us to expression 
\begin{equation}
S\left(t\right)=\frac{1}{2}\left\{ {\rm erf}\left[\frac{z_{0}}{\sqrt{2}\sigma}\right]+{\rm erf}\left[\frac{Ze^{-\alpha t}-z_{0}}{\sqrt{2}\sigma}\right]\right\} .\label{eq: Final expression for the survival probability}
\end{equation}
We also note, that 
\[
\intop_{0}^{+\infty}dzP\left(z,t\right)<1,
\]
since the animals die and hence disappear at the infinity.

\subsection{Estimating the first passage time, the survival probabilities, and
the average lifespan\label{subsec:Estimating-the-FPT}}

In the case under consideration, a more interesting quantity to calculate
and analyze is not the probability density $P(z,t)$ itself, but a
so-called first passage time distribution function $P_{{\rm FPT}}(t)$
\cite{Redner2001,Aalen2008}. The reason is that the probability density
$P(z,t)$ represents a sum over all possible stochastic trajectories
$z(t)$, including those which pass the effective threshold $z=Z$
and then return back. On the other hand, by definition, $P_{{\rm FPT}}(T)$
is a probability for a given stochastic trajectory $z=z(t)$ to pass
the threshold $z=Z$ at $t=T$ for the very first time. As such, it
does not take into account trajectories which then re-enter the allowed
domain of $z$.

\begin{figure}
\includegraphics[width=0.5\textwidth]{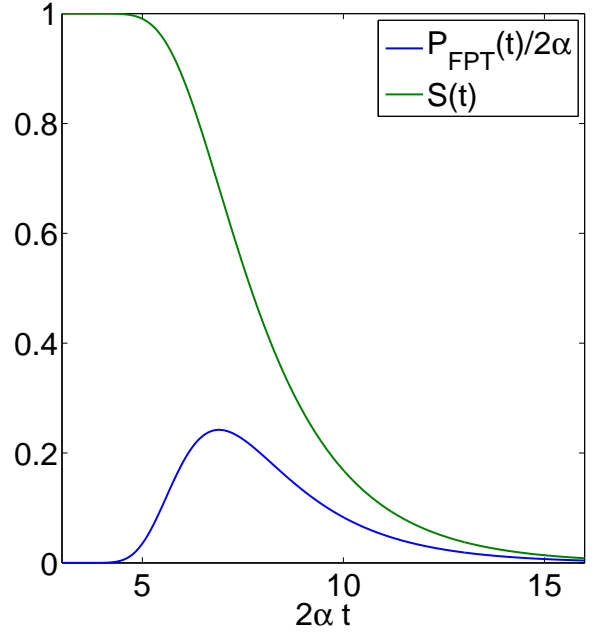}\caption{The first passage time distribution function $P_{{\rm FPT}}(t)$ (in
units of $2\alpha$) and the survival probability $S(t)$ as a function
of time $2\alpha t$; $\frac{\alpha Z^{2}}{\Delta}=1000.$\label{fig:The-first-passage}}
\end{figure}

When the ``deterministic'' form of the initial condition is chosen,
the first passage time distribution function is given by
\[
P_{{\rm FPT}}(t)=-\dot{S}\left(t\right)=
\]
\begin{equation}
=\sqrt{\frac{\alpha}{\pi\varDelta}}\frac{2\alpha Ze^{2\alpha t}}{\left[X\left(t\right)\right]^{3/2}}\exp\left[-\frac{\alpha Z^{2}}{X\left(t\right)\varDelta}\right].\label{eq:FPTexact}
\end{equation}
The function (\ref{eq:FPTexact}) grows exponentially at small $t\lesssim\alpha^{-1}$,
possesses a distinct maximum at $t_{{\rm max}}=t_{ls}=\alpha^{-1}\ln\left[\left(\frac{\alpha Z^{2}}{\Delta}\right)^{1/2}\right]$,
with its width $\delta t\sim\alpha^{-1}$ near the maximum and finally
decays as $e^{-\alpha t}$ at $t\gg t_{{\rm max}},$ see Fig. \ref{fig:The-first-passage}.
Computing the first passage time distribution function also allows
one to easily estimate the survival probability $S(t)$, using the
prescription $P_{FPT}(t)=-dS(t)/dt$. From Eq. (\ref{eq:FPTexact})
we find 
\begin{equation}
S(t)={\rm erf}\left[\sqrt{\frac{\alpha Z^{2}}{2\Delta X(t)}}\right],\label{eq:SurvivalProbabilityLate-2}
\end{equation}
In the regime $\alpha^{-1}<t\lesssim t_{ls}$ , also known as Suzuki
scaling \cite{Suzuki1977,Suzuki2009},equation for $P_{{\rm FPT}}(t)$
reduces to
\begin{equation}
P_{{\rm FPT}}(t)\approx2\alpha Z\left(\frac{\alpha}{2\pi\Delta}\right)^{1/2}\exp\left(-\alpha t-\frac{\alpha Z^{2}}{2\Delta}e^{-2\alpha t}\right).\label{eq:FPTprobSuzuki}
\end{equation}

If the non-linearity is small, the center of the originally narrow
distribution in the $z$-space follows the well defined trajectory
$\bar{z}\left(t\right)=z_{0}\exp\left(\alpha t\right)$. At $t=t_{0}=\alpha^{-1}\ln\left(Z/z_{0}\right)$
it achieves the absorbing wall, which means that the average lifespan
in the population described by the distribution is also $t_{ls}\approx t_{0}$.
This result follows immediately from (\ref{eq: Final expression for the survival probability})
and the exact definition of the average lifespan
\begin{equation}
t_{ls}=\intop_{0}^{\infty}dtS\left(t\right).\label{eq: Average lifespan}
\end{equation}

From Eq. (\ref{eq:FPTprobSuzuki}) we find
\begin{equation}
S(t)\approx{\rm erf}\left[\left(\frac{\alpha Z^{2}}{2\Delta}\right)^{1/2}e^{-\alpha t}\right],\label{eq:SurvivalProbabilityLate}
\end{equation}
expression precise up to terms, exponentially small at $\frac{\alpha Z^{2}}{\Delta}\gg1$,
i.e., when a species is considered long-lived compared to the characteristic
time scales of molecular dynamics in the cell, see the next Subsection.
The function (\ref{eq:SurvivalProbabilityLate}), initially exponentially
close to $1$, then falls off quickly (as we shall see below, this
fall-off corresponds to an exponentially growing mortality rate) and
finally approaches zero as $\exp(-\alpha t)$ at $t\gg t_{ls}$.

\subsection{Behavior of the mortality rate}

As usual, the mortality rate is defined as $M(t)=-S^{-1}\cdot dS/dt$,
so that one has precisely
\begin{equation}
M(t)=\frac{P_{{\rm FPT}}(t)}{S(t)}.\label{eq:MortalityRateDef}
\end{equation}
Once again, for the very narrow initial distribution (\ref{eq:BoundaryConditionNlin}),
$\sigma_{0}\rightarrow0$ and $z_{0}=0$ ($G(z)\approx\delta(z)$),
we find, that
\[
M\left(t\right)=\sqrt{\frac{\alpha}{\pi\varDelta}}\frac{2\alpha Ze^{2\alpha t}}{\left[X\left(t\right)\right]^{3/2}}\exp\left[-\frac{\alpha Z^{2}}{X\left(t\right)\varDelta}\right]\times
\]
\begin{equation}
\times\left\{ {\rm erf}\left[\sqrt{\frac{\alpha Z^{2}}{2X(t)\Delta}}\right]\right\} .^{-1}\label{eq: Mortality for the deterministic case}
\end{equation}
For a more realistic determenistic form of the probability distribution
at the origin \ref{eq: Motion only to the right} the solution is
still possible, and for sufficiently advanced ages, $t\gg\alpha^{-1}$,
we obtain
\[
M\left(t\right)=\frac{Z\alpha}{\sigma}\sqrt{\frac{2}{\pi}}\exp\left[-\frac{\left(Ze^{-\alpha t}-z_{0}\right)^{2}}{2\sigma^{2}}-\alpha t\right]\times
\]
\begin{equation}
\times\left\{ 1+{\rm erf}\left[\frac{Ze^{-\alpha t}-z_{0}}{\sqrt{2}\sigma}\right]\right\} ^{-1},\label{eq: Mortality for the Gaussian initial distribution}
\end{equation}
once the non-linearity is sufficiently weak, as commanded by the condition
(\ref{eq: Main large parameter}).

\subsubsection{Vanishing mortality at very early ages $t\ll\alpha^{-1}$}

At sufficiently early ages $t\ll\alpha^{-1}$ behavior of $z$ is
dominated by the effect of the random force $f(t)$ over the ``deterministic''
force $v(z)$ \cite{Suzuki1977,Suzuki2009}. Accordingly, the state
variable and its fluctuations grow exponentially with age. This conclusion
is confirmed by the analysis of senescence-associated transcriptional
dynamics. It is worth emphasizing that the characteristic time scale
of the stochastic growth of the expressome is $\sim\alpha^{-1}$.
It thus depends only on the morphological properties of the GRN and
does not depend on the genotoxic stress amplitude $\Delta$ at all. 

At very early ages $t\ll\alpha^{-1}$ the first passage time distribution
function is exponentially strongly suppressed:

\begin{equation}
P_{{\rm FPT}}(t)\approx\frac{2\alpha Z}{(2\alpha t)^{3/2}}\left(\frac{\alpha}{2\pi\Delta}\right)^{1/2}\exp\left(-\frac{Z^{2}}{4\Delta t}\right).\label{eq:VeryEarlyMortality}
\end{equation}
In this regime the survival probability $S(t)$ remains exponentially
close to $1$:

\[
S(t)\approx1-\frac{2}{\sqrt{\pi}}\left(\frac{4\Delta t}{Z^{2}}\right)^{2}\exp\left(-\frac{Z^{2}}{4\Delta t}\right).
\]
Therefore, the mortality rate remains exponentially close to $0$,
and one approximately has $M(t)\approx P_{{\rm FPT}}(t)$. In our
opinion, this behavior might explain relatively small mortality, often
observed in cohorts at very early ages. We note that the behavior
of the mortality rate in this regime is not universal as it strongly
depends on the amplitude $\Delta$ of stochastic genotoxic stress.

\subsubsection{Gompertz behavior at later ages $t\sim\alpha^{-1}$}

As was explained above, as the age increases, the effect of the stochastic
force $\delta F$ in Eq. (\ref{eq: Main equation}) quickly becomes
negligible in comparison to the effect of the ``deterministic''
force $-K\delta x$; the stochastic regime is changed by the drift
motion at $t\sim\alpha^{-1}$. Correspondingly, the behavior (\ref{eq:VeryEarlyMortality})
quickly changes when $t\agt\alpha^{-1}$, yielding a Gompertz-like
dependence of the mortality rate on age: $M\left(t\right)\approx M_{0}\exp\left(\gamma t\right)$,
where the exponent 
\[
\gamma\left(t\right)=\frac{d}{dt}\ln M\left(t\right)\sim\frac{\alpha Z^{2}\dot{X}\left(t\right)}{X^{2}\left(t\right)\varDelta}\sim\alpha\frac{\alpha Z^{2}}{\varDelta}\gg\alpha
\]
is found from Eq. (\ref{eq: Mortality for the deterministic case})
and is much greater than the GNR instability parameter $\alpha$.

\subsubsection{Universal Gompertz behavior in the regime of Suzuki scaling}

At even later times of the order $t\sim t_{ls}$, when the regime
of Suzuki scaling is realized, we also find a universal Gompertz behavior.
As was explained above, the first passage time probability distribution
$P_{{\rm FPT}}$ reaches a maximum at $t_{{\rm max}}=\alpha^{-1}\ln\Lambda=t_{ls}$,
where $\Lambda=\frac{\alpha^{1/2}Z}{\Delta^{1/2}}$, and its half-width
near the the maximum is of the order $\alpha^{-1}$, as in Fig. \ref{fig:The-first-passage}.
By the time $t\sim t_{{\rm max}}$ the argument of the error function
in Eq. (\ref{eq:SurvivalProbabilityLate}) is already sufficiently
small for it to be expanded. One thus has

\[
S(t)\approx\frac{2}{\sqrt{\pi}}\Lambda e^{-\alpha t}-\frac{2}{3\sqrt{\pi}}\Lambda^{3}\exp(-3\alpha t)+\ldots.
\]
On the other hand, one finds for $P_{{\rm FPT}}(t)$ in the same regime
\[
P_{{\rm FPT}}(t)=\frac{2\alpha\Lambda}{\sqrt{\pi}}e^{-\Lambda^{2}\exp(-2\alpha t)}e^{-\alpha t},
\]
so that approximately 
\[
M(t)\approx\alpha e^{-\Lambda^{2}\exp(-2\alpha t)}.
\]
Introducing $t=t_{{\rm max}}+\delta t$ one finally finds at small
$\delta t\lesssim(2\alpha)^{-1}$
\begin{equation}
M(t)\approx e^{-1/2}\alpha e^{\alpha\delta t}\approx0.61\alpha e^{\alpha\delta t}.\label{eq:GompertzUniversal}
\end{equation}
Thus, the Gompertz law with the universal exponent $\alpha^{-1}$
is realized at times $t_{{\rm max}}-(2\alpha)^{-1}\lesssim t\lesssim t_{{\rm max}}+(2\alpha)^{-1}$.
We again emphasize that the Gompertz aging rate $\alpha$ depends
only on the properties of the GRN under consideration and does not
depend on stress. In fact, the very same conclusion holds for the
Gaussian initial distribution case (\ref{eq: Mortality for the Gaussian initial distribution}).

\subsubsection{Mortality deceleration at late ages\label{subsec:Slowing-down-of-mortality}}

\begin{figure}
\includegraphics[width=0.95\columnwidth]{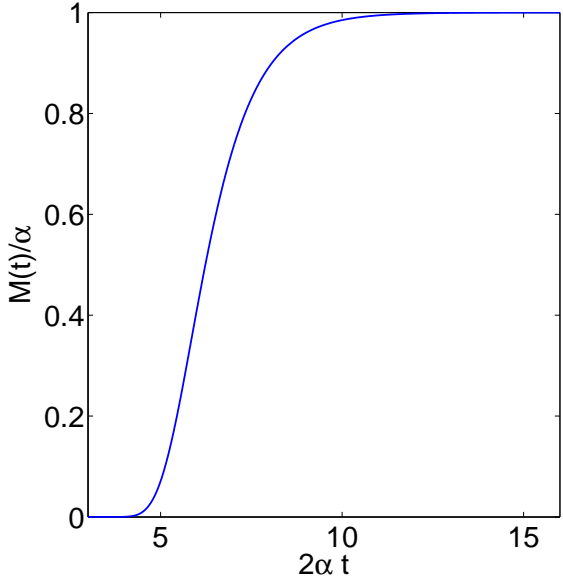}\caption{The mortality rate $M(t)$ for the case of the ``diffusive'' initial
condition; $\alpha Z^{2}/\varDelta=1000.$ The maximum of $P_{{\rm FPT}}(t)$
is reached at $2\alpha t\approx6.9$.\label{fig:The-mortality-rate}}
\end{figure}

Finally, at $t\gg t_{{\rm max}}$ both the first passage time distribution
function (\ref{eq:FPTexact}) and the survival probability (\ref{eq:SurvivalProbabilityLate-2})
decay exponentially. This corresponds to an asymptotically constant
behavior of the mortality rate in both cases, (\ref{eq: Mortality for the deterministic case})
and (\ref{eq: Mortality for the Gaussian initial distribution}):
\begin{equation}
M(t)\approx\alpha,\label{eq:MortalitySlowdown}
\end{equation}
see Fig. \ref{fig:The-mortality-rate}. For late $t$, corrections
to the leading constant term become exponentially small with time.
The existence of mortality plateau in models of the type (\ref{eq: Main equation})
is known in literature \cite{Weitz2001,Steinsaltz2004}.

\begin{figure}
\includegraphics[width=0.95\columnwidth]{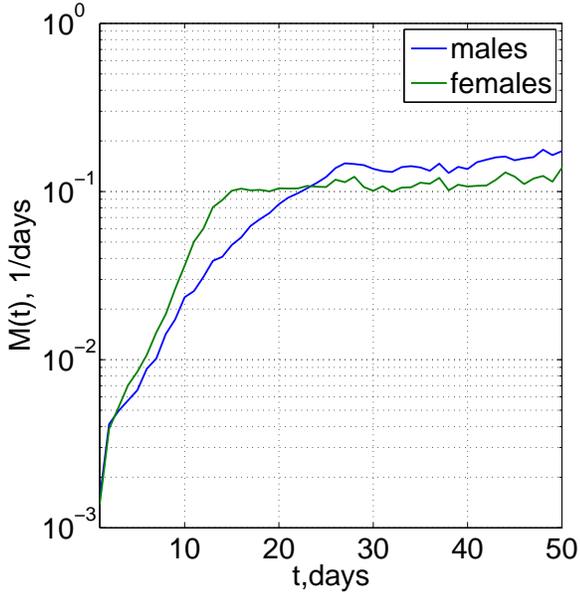}\caption{The mortality rate as a function of age from the very large cohorts
study, medflies, males and females from \cite{vaupel1998biodemographic}.\label{fig:The-mortality-rate-medflies}}
\end{figure}

\subsection{Concluding remarks on aging regimes phenomenology\label{subsec:Phenomenology}}

\subsubsection{The unstable case, Gompertz limit, $\alpha Z^{2}/\varDelta\gg1$}

This case includes situations with a very weak non-linearity in the
Fokker-Planck equation (\ref{eq:FokkerPlanck}) and corresponds to
a very high toxicity threshold $Z$ as compared to the characteristic
amplitude of fluctuations of the stress factors $\Delta.$ As was
explained above, we expect a species with such a GRN to be relatively
long-lived, as the average life expectancy (\ref{eq: Age of achieving the absorbing wall})
is logarithmically larger than the inverse Gompertz exponent $\alpha^{-1}$.
Four very distinct regimes of mortality exist for such species: (a)
the regime of small mortality at early ages, (b) a Gompertz regime
with the Gompertz exponent $\sim\alpha^{2}Z^{2}/\varDelta\gg\alpha$,
(c) a Gompertz regime with a universal Gompertz exponent $\alpha$
and finally (d) a regime of constant mortality $M_{\infty}=\alpha$.
On the other hand, if indeed $Z\gg\sigma_{0}$, the initial Gompertz
exponent in regime (b) is very large, and a population of moderate
size simply dies out before the regime of mortality slowing-down is
observed. 

\subsubsection{The unstable case, short lifespan limit, $\alpha Z^{2}/\varDelta\gtrsim1$}

There are again four distinct regimes (a-d) of aging discussed above,
with that difference that the regime (b) is hard to observe or unobservable,
so a Gompertz regime (c) with the exponent $\sim\alpha$ is observed.
Clearly, the regimes (c) and (d) are closely connected to each other
in the sense, described above, and there is a strong correlation between
the observed value of the Gompertz exponent and the mortality rate
$M_{\infty}$ at late ages. This, as it seems, is what we observe
e.g. in experiments with extremely large cohorts study of medflies,
males and females from \cite{vaupel1998biodemographic}, see Fig.
\ref{fig:The-mortality-rate-medflies}. A relatively short average
lifespan of such species is comparable to the value of Gompertz exponent.

\subsubsection{The stable case, $\alpha<0$\label{subsec:NS-case-Mortality}}

In this case, arguably the most interesting to consider, the effective
potential $V(z)$ of the Langevin equation possesses a local minimum,
separated from the roll-down to $z\to\infty$ by a potential barrier,
see Fig. \ref{fig:SchematicVandPCA}B of the main text. There exists
a single regime of aging realized for the GRN of this type. It is
characterized by a constant mortality rate $M_{\infty}$, its value
being determined by the exponentially small Kramers probability of
passage through the barrier\cite{lifshitz1995physical}. If a GRN
with such properties indeed exists in nature, it would correspond
to a species with negligible senescence. We leave the in-depth study
of this regime for a future work. 

\section{Gene annotation}

To analyze the vector $b$, obtained from the transcriptome of \emph{D.
melanogaster} \cite{pletcher2002genome}, we have performed a gene
set enrichment analysis for the components of vector $b$ using \textquotedblleft Parent-Child-Intersection\textquotedblright{}
GO enrichment analysis procedure described in \cite{grossmann2007improved}.
This method considers genes in the background if they are present
in all parent terms of the term of interest. Since the values in vector
$b$ can be both positive and negative, lists of positive and negative
leading components were analyzed separately. We studied the dependence
of significantly enriched GO categories on the number of the leading
components selected. The list of enriched GO categories appeared to
be stable in a broad range of number of the leading components, so
200 positive and 200 negative leading components from GO-stability
regions were analyzed.We employed an adjusted $p$-value (Benjamini-Hochberg)
cutoff of $0.05$.

We have performed KEGG pathway enrichment analysis of the human orthologs
of the genes corresponding to the most significant components of vector
$b$, computed from the transcriptome of \emph{D. melanogaster }\cite{pletcher2002genome}.
A one-sided Fischer exact test was performed, after which an adjusted
$p$-value (Benjamini-Hochberg) cutoff of $0.05$ was employed. We
used OrthoDB v.7 \cite{waterhouse2013orthodb} to map fly genes to
corresponding human orthologs.

\bibliographystyle{naturemag}
\bibliography{Qrefs}

[Competing Interests] The authors declare that they have no competing
financial interests. 

[Author Contrbutions] All authors contributed equally to this work.

[Correspondence] Correspondence and requests for materials should
be addressed to P.O.F.~(email: peter.fedichev@gmail.com)
\end{document}